\documentclass[twocolumn]{aastex701}
\usepackage{amsmath}
\usepackage{natbib}
\usepackage{xcolor}
\usepackage{todonotes}

\newcommand{\mbh}{M_{\rm BH}}
\newcommand{\mbhzero}{M_{\rm BH,0}}
\newcommand{\mdot}{\dot{M}_{0}}
\newcommand{\msun}{M_{\odot}}
\newcommand{\astar}{a_{\rm \star}}
\newcommand{\abhzero}{a_{\rm \star,0}}
\newcommand{\aeq}{a_{\star; \rm eq}}
\newcommand{\afinal}{a_{\star; \rm f}}
\newcommand{\rg}{r_{\rm g}}
\newcommand{\rgzero}{r_{\rm g, 0}}
\newcommand{\phit}{\tilde{\phi}}
\shorttitle{Black Hole Spin Evolution in Magnetized Accretion Flows}
\shortauthors{D. Issa, O. Gottlieb, J. Jacquemin-Ide \& M. Liska}
\begin{document}

\title{Black Hole Spin Evolution in Magnetized Accretion Flows}

\author[orcid=0009-0005-2478-7631,sname='Danat Issa']{Danat Issa}
\affil{Department of Physics and Kavli Institute for Astrophysics and Space Research, Massachusetts Institute of Technology, Cambridge, MA 02139, USA}
\email[show]{danat@mit.edu}

\author[orcid=0000-0003-3115-2456]{Ore Gottlieb}
\affil{Department of Physics and Kavli Institute for Astrophysics and Space Research, Massachusetts Institute of Technology, Cambridge, MA 02139, USA}
\email{ogot@mit.edu}

\author[orcid=0000-0003-2982-0005]{Jonatan Jacquemin-Ide}
\affiliation{JILA, University of Colorado and National Institute of Standards and Technology, 440 UCB, Boulder, CO 80309-0440, USA}
\email{Jonatan.Jacquemin@colorado.edu}

\author[orcid=0000-0003-4475-9345]{Matthew Liska}
\affiliation{Center for Relativistic Astrophysics, Georgia Institute of Technology, Howey Physics Bldg, 837 State St NW, Atlanta, GA 30332, USA}
\email{mliska3@gatech.edu}

\begin{abstract}
Black hole (BH) spin grows as accretion supplies angular momentum and decreases as magnetically driven jets remove it, driving the BH toward an equilibrium spin, $a_{\rm \star; eq}$. Existing estimates of $a_{\rm \star; eq}$ rely on general-relativistic magnetohydrodynamic (GRMHD) simulations with fixed BH mass and spin, and cover only the extreme limits of horizon magnetic flux $\phi\rightarrow0$ and magnetically arrested disks (MADs), $\phi\approx\phi_{\rm MAD}$. We evolve the BH mass and spin in long 3D GRMHD simulations of radiatively inefficient accretion flows aligned or anti-aligned with the BH spin. We find the spin-up parameter as a function of BH spin and $\phi/\phi_{\rm MAD}$, and derive the equilibrium spin as a continuous function of flux, $a_{\rm \star; eq}(\phi/\phi_{\rm MAD})$. The equilibrium spin falls steeply with flux, from $a_{\rm \star; eq}\approx0.9$ in weakly magnetized flows through $a_{\rm \star; eq}(\phi/\phi_{\rm MAD}=0.5)\approx0.5$ to $a_{\rm \star; eq}(\phi/\phi_{\rm MAD}=1)\approx0.06$ in sustained MADs. In weakly magnetized flows, a slowly spinning BH must accrete at least its initial mass to reach equilibrium, whereas MADs spin down rapidly rotating BHs after accreting less than half of it. However, as the BH grows, the growing gravitational radius and enhanced inflow reduce $\phi/\phi_{\rm MAD}$, driving disks whose magnetic-flux supply cannot keep pace out of the MAD state and allowing the BH to transition from spin-down to spin-up. The final spin thus depends on both the magnetic flux history and the accreted mass. We discuss implications for a variety of BH-powered engines and provide a subgrid prescription for galaxy-scale simulations and population studies.
\end{abstract}

\section{Introduction} \label{sec:introduction}

Accreting black holes (BHs) power some of the most luminous sources in the universe, from X-ray binaries and active galactic nuclei (AGN) to gamma-ray bursts \citep[GRBs;][]{meier_bh_engine_2012}. The BH spin shapes the dynamics of the inner accretion flow and affects the efficiency with which accretion converts rest-mass energy into radiation. Magnetic fields threading the BH horizon can extract the BH rotational energy to power relativistic jets through the Blandford--Znajek mechanism \citep[BZ;][]{blandford_electromagnetic_1977}. Therefore, BH spin plays a central role in setting the radiative and mechanical feedback into the surrounding medium, linking BH physics to the evolution of astrophysical systems.

The BH spin evolves through the exchange of energy and angular momentum with its surroundings. Accretion of angular momentum from the disk can spin up the BH, while magnetically driven BZ jets exert a spin-down torque. The competition between these processes drives the BH toward an equilibrium spin, $\aeq$, whose value depends on the properties of the accretion flow and the magnetic flux threading the horizon \citep{gammie_black_2004}. However, whether the BH approaches this equilibrium depends on the accreted mass relative to the BH mass, which may vary across astrophysical systems.

Several observational techniques constrain BH spins. Gravitational-wave (GW) observations of merging binary BHs favor predominantly slow-to-moderate spins, with GWTC-4 population analyses inferring that $\sim90\%$ of BHs in merging binaries have spin magnitudes $\chi < 0.6$ \citep{abbott_gwtc-4_2025}. Recent population analyses also find evidence for mass-dependent spin distributions, including broader spin distributions at higher masses \citep[e.g.,][]{Antonini2025,Banagiri2026,Flanagan2026,Guttman2026,hussain_evidence_2026,Ray2026,Wolfe2026}. GW measurements probe BHs at the end of their binary evolution, where the BH spin may reflect stellar evolution or previous mergers, making it difficult to isolate the contribution of accretion to their spin evolution \citep[see][for a review]{biscoveanu_first_2026}.

Spins of accreting BHs are inferred from the emission of the inner accretion flow \citep{Reynolds2021}. X-ray reflection spectroscopy \citep{garcia_2014} favors predominantly high spins in X-ray binaries (XRBs; \citealt{draghis_systematically_2024}), whereas thermal continuum fitting \citep{zhang_1997, zhu_2012} yields a wider range \citep{mcclintock_2014}. AGN estimates also span a range of values, with selection effects favoring rapidly spinning BHs \citep{sisk-reynes_spin_2026}. Both methods rely on assumptions about the inner accretion flow whose systematic uncertainties remain poorly understood \citep{shashank_measuring_2025, tausch_spinning_2026}. This may contribute to the apparent difference between XRB and GW spin distributions \citep{zdziarski_spins_2026}, which has alternatively been interpreted as evidence that the two techniques sample distinct BH populations \citep{Fishbach2022,gallegos-garcia_high-spin_2022}. Interpreting the tension between these measurements requires understanding how accretion modifies the BH spin over time.

Equilibrium spins are generally inferred from general-relativistic magnetohydrodynamic (GRMHD) simulations with fixed BH mass and spin. A standard, razor-thin, radiatively efficient Novikov--Thorne \citep[NT;][]{novikov_astrophysics_1973} disk with a Keplerian rotation profile spins the BH up to $\aeq = 1$ \citep{bardeen_kerr_1970}, or to $\aeq \approx 0.998$ once the capture of disk photons by the BH is included \citep{thorne_disk-accretion_1974}. Reaching this limit requires substantial accretion: a non-spinning BH must grow its mass by a factor of $\sqrt{6} \approx 2.45$ to reach $\astar = 1$ \citep{bardeen_kerr_1970}. Hot, advection-dominated accretion flows (ADAFs) also drive the BH to an equilibrium spin $\aeq\approx0.7$--$1$, depending on the efficiency of angular momentum transport \citep{Popham1998}. Calibrating the transport with MHD simulations gives $\aeq \approx 0.96$, which jet torques lower to $\approx0.93$ \citep{Benson2009}. GRMHD simulations of weakly magnetized SANE \citep[standard and normal evolution;][]{narayan_2012} flows show that they accrete near-thin-disk specific angular momentum and energy and launch weak jets, resulting in similarly high $\aeq \approx 0.9$--$0.94$ \citep{gammie_black_2004,chatterjee_mad_eruptions_2022,dhruv_survey_2025}.

In the opposite limit of a magnetically arrested disk \citep[MAD;][]{bisnovatyi-kogan_mad_1974,narayan_mad_2003,igumenshchev_2003,igumenshchev_magnetically_2008,tchekhovskoy_efficient_2011,Tchekhovskoy2012}, large-scale magnetic flux becomes dynamically important near the BH. The disk becomes sub-Keplerian, and the jets remove more angular momentum from the BH than the disk supplies, rapidly spinning down BHs with $\astar > \aeq$ \citep{Tchekhovskoy2012,narayan_jets_2022, lowell_rapid_2024, dhruv_survey_2025}. Global GRMHD simulations of radiatively inefficient (hot) MADs find $\aeq \approx 0.035$--$0.07$ \citep{Tchekhovskoy2012,narayan_jets_2022,lowell_rapid_2024}, and a BH needs to accrete only $\sim 20\%$ of its mass to spin down from $\astar = 1$ to $0.2$ \citep{lowell_rapid_2024}. Radiative MADs near the Eddington rate were found to reach $\aeq \approx 0.8$ \citep{ricarte_recipes_2023}, whereas thin, luminous MADs settle at $\aeq \approx 0.3$  \citep{lowell_evidence_2025} and neutrino-cooled collapsar disks at $\aeq \approx 0.13$ \citep{issa_collapsar_2026}. However, the near-Eddington disks of \citet{ricarte_recipes_2023} hold only about a third of the saturated magnetic flux, suggesting that the difference may stem from the magnetic flux on the BH rather than from cooling alone. Therefore, interpreting these results requires mapping $\aeq$ across the full range of magnetic flux.

The equilibrium spin at intermediate magnetic flux remains unknown. Although fixed-metric simulations can be used to infer spin evolution, they cannot capture the feedback of the evolving BH mass and spin on the accretion flow, particularly when the magnetic state changes during accretion. Over the past decade, galaxy-scale simulations have increasingly evolved BH spin on the fly \citep{dubois_2014, dubois_2021, talbot_2021, husko_2022}, using spin-up prescriptions for either thin disks or fully MAD flows \citep[e.g.,][]{massonneau_2023}. However, a prescription that follows the spin evolution at intermediate horizon magnetic flux, between the SANE and MAD limits, is still missing.

In this paper, we present an implementation of BH mass and spin evolution in GRMHD. We perform a suite of 3D GRMHD simulations of radiatively inefficient, magnetized accretion disks, varying the torus-to-BH mass ratio and the initial poloidal magnetic field configuration to explore different BH growth rates and accretion states, ranging from SANE to MAD. Using these simulations, we fit the BH spin evolution as a function of instantaneous spin and horizon magnetic flux, and determine the equilibrium spin as a function of magnetic flux. We begin by presenting our results for the general spin evolution in Section~\ref{sec:spin_evolution}. We describe the implementation in Section~\ref{sec:methods} and the numerical simulations in Section~\ref{sec:simulations}. We conclude, discuss the astrophysical implications, and provide a subgrid prescription for cosmological simulations in Section~\ref{sec:discussion}.

\section{Black hole spin evolution in a nutshell}\label{sec:spin_evolution}

In this section, we summarize our main results, which are described in detail in Sec.~\ref{sec:results}.

\subsection{Equilibrium spin}

To find the equilibrium spin of the BH, $\aeq$, we consider the spin-up parameter \citep{gammie_black_2004},
\begin{equation} \label{eq:spin-up-parameter}
    s \equiv \frac{\mbh}{\mdot} \frac{d\astar}{dt},
\end{equation}
where $\mbh$ is the BH mass, $\mdot$ is the rest-mass accretion rate, and $\astar$ is the instantaneous BH spin. The spin-up parameter is positive (negative) when the BH spins up (down). The equilibrium spin is the final spin attained by the BH in the limit of infinite mass accretion. Therefore, it does not depend on the mass accretion rate, but only on the dimensionless magnetic flux on the BH,
\begin{equation}\label{eq:phi0}
    \phi = \frac{\Phi_{\rm H}}{\sqrt{\mdot\rg^2c}},
\end{equation}
where $\Phi_{\rm H}$ is the poloidal magnetic flux threading the BH horizon [see Eq.~\eqref{eq:phi}], and $\rg = G\mbh/c^2$ is the gravitational radius of the instantaneous BH mass. We present our results as a function of the normalized dimensionless magnetic flux, known as the ``MAD'' parameter, $\phit \equiv \phi/\phi_{\rm MAD}$, which varies from $\phit = 0$ in the absence of magnetic fields to $\phit \approx 1$ at the magnetic flux saturation value \citep{narayan_jets_2022},
\begin{equation}
\phi_{\rm MAD}(\astar) = 52.6 + 34 \astar - 14.9 \astar^2 - 20.2 \astar^3.
\end{equation}

\begin{figure}
    \centering
    \includegraphics[width=\linewidth]{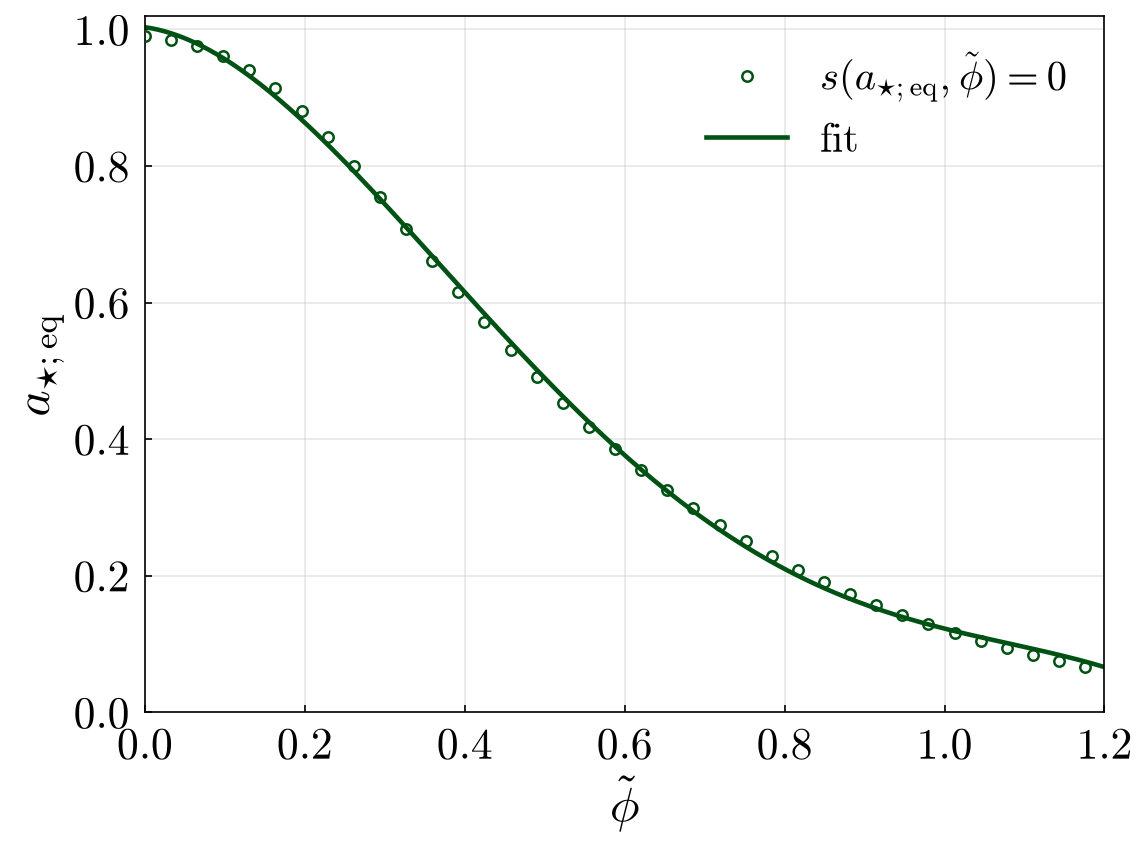}
    \caption{Equilibrium spin as a function of the MAD parameter. The circles are roots of $s(\aeq, \phit) = 0$, and the line marks the fit in Eq.~\eqref{eq:aeq_fit}.}
    \label{fig:aeq-fit}
\end{figure}

Figure~\ref{fig:aeq-fit} shows the equilibrium spin obtained by solving $s(\aeq, \phit) = 0$. We approximate this relation with a fourth-order polynomial,
\begin{equation} \label{eq:aeq_fit}
    \aeq (\phit) = 1 - 1.47\, \phit^4 + 4.27\, \phit^3 - 3.55\, \phit^2 - 0.13\, \phit.
\end{equation}
The $\phit$ axis extends beyond unity because $\phi_{\rm MAD}$ is a time-averaged saturation level: flux eruptions make $\phi$ fluctuate by tens of percent around it, introducing a large uncertainty in the MAD regime.

At $\phit = 0$, the NT limit is approximately recovered, and $\aeq$ decreases monotonically with $\phit$. At $\phit = 1$, however, the fit gives $\aeq \approx 0.12$. This value is $2$--$3$ times higher than those found in previous MAD studies \citep{narayan_jets_2022,lowell_rapid_2024}. This discrepancy arises from two effects: (i) near $\astar \approx 0.05$, our MADs saturate at $\phit\approx1.1$--$1.2$ (see Figure~\ref{fig:phi-mad-fit}), so that sustained MADs sit at $\phit\gtrsim1$, where Eq.~\eqref{eq:aeq_fit} gives $\aeq\approx0.06$--$0.09$; (ii) the assumed linear relation between the hydrodynamic component of the specific angular momentum, $l_{\rm HD}$, and $\phit$ overestimates $l_{\rm HD}$ in MADs near $\astar \approx 0$. Adopting instead the numerical value $l_{\rm HD}/l_{\rm NT} = 0.15$--$0.2$ at $\phit \approx 1$ [Figure~\ref{fig:specific-fluxes-fit}(d)] gives $\aeq \approx 0.050$--$0.067$.\footnote{Our simulations, like those of \citet{lowell_rapid_2024}, use $\gamma_{\rm ad} = 4/3$, whereas \citet{narayan_jets_2022} used $13/9$. The adiabatic index does not affect the saturated flux \citep{narayan_jets_2022}, which justifies adopting their $\phi_{\rm MAD}(\astar)$ for our simulations. However, it does affect the spin-up: \citet{lowell_rapid_2024} found that $\gamma_{\rm ad} = 13/9$ yields systematically lower $s$, and attributed part of the difference between their $\aeq \approx 0.07$ and the lower $\aeq$ of \citet{narayan_jets_2022} to the adiabatic index. Our $\aeq \approx 0.05$--$0.067$ lies between these values.}

\subsection{Final spin}

\begin{figure*}
    \centering
    \includegraphics[width=\linewidth]{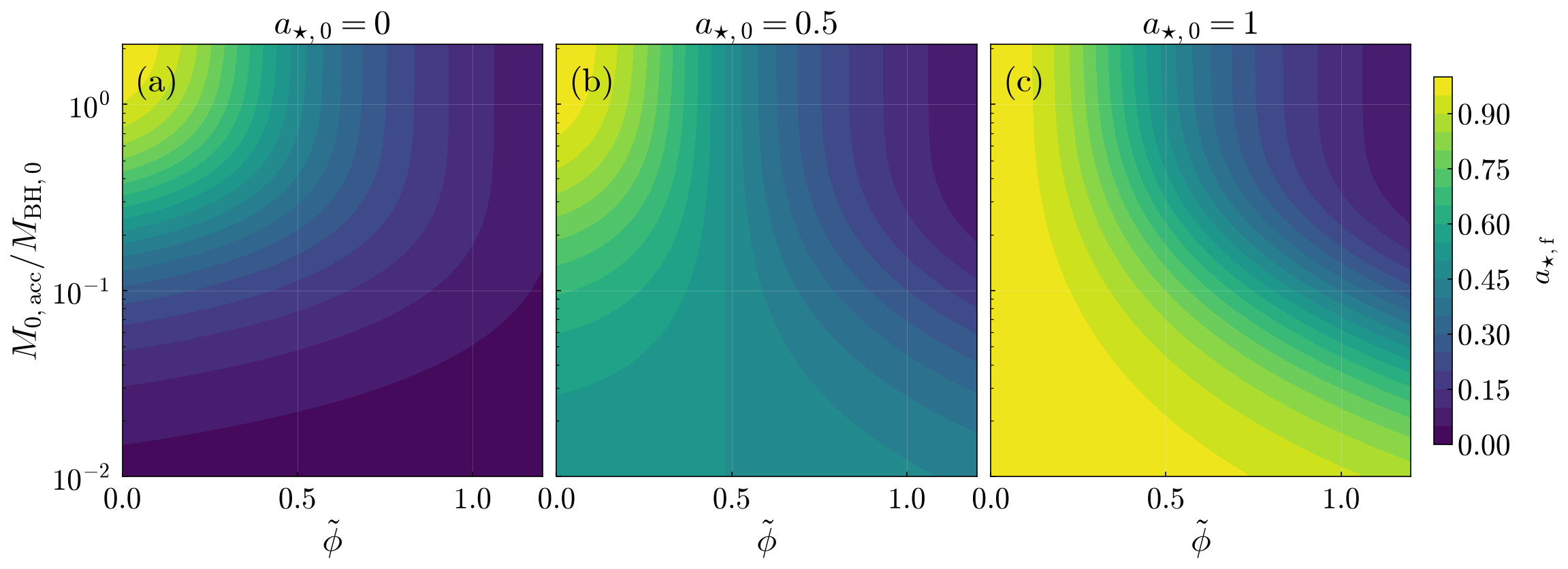}
    \caption{Final BH spin, $\afinal$, as a function of (fixed) $\phit$ and of the ratio of the accreted rest mass, $M_{0,\,\rm acc}$, to the initial BH mass, shown for initial spins $\abhzero = 0$ (a), $\abhzero = 0.5$ (b), and $\abhzero = 1$ (c).}
    \label{fig:af-vs-m-phi}
\end{figure*}

The equilibrium spin, $\aeq$, is the asymptotic value that the spin reaches if $\phit$ stays fixed and the mass reservoir is large enough. However, a finite-mass reservoir leaves the BH with a different final spin, $\afinal$. In general, both $\mdot$ and $\phit$ evolve in time, as in our simulations, and one needs to integrate the fits along $\mdot(t)$ and $\phit(t)$. Because $s$ depends nonlinearly on $\phit$, integrating the fits along $\phit(t)$ is not equivalent to evaluating them at the time-averaged $\phit$. For the purpose of estimating the final spin, we assume that both $\mdot$ and $\phit$ are constant in time. Using Eq.~\eqref{eq:spin-up-parameter} and the accreted rest mass, $dm^\prime = \mdot\, dt$, we write the change in the BH spin as
\begin{equation} \label{eq:delta_a}
    \Delta \astar = \int_{0}^{M_{\rm 0, acc}} \frac{s (\astar, \phit)}{\mbh} dm^\prime,
\end{equation}
where the BH mass evolves with the accreted rest mass as
\begin{equation} \label{eq:dMdm}
    \frac{d\mbh}{dm^\prime} = e_{\rm HD}(\astar, \phit) + e_{\rm EM}(\astar, \phit).
\end{equation}
Here, $e_{\rm HD}$ and $e_{\rm EM}$ are the hydrodynamic and electromagnetic components of the specific energy flux through the horizon [see Eq.~\eqref{eq:l_and_e_def}]. We integrate Eqs.~\eqref{eq:delta_a} and \eqref{eq:dMdm} from the initial spin, $\abhzero$, and initial mass, $\mbhzero$, using the fits to our simulations (see Table~\ref{tab:fits}). Because $s$ and $e$ depend only on $\astar$ and $\phit$, for a given $\abhzero$ and $\phit$ the final spin, $\afinal$, depends only on the accreted rest mass, $M_{\rm 0,acc}=\int\mdot\,dt$, and not on the accretion rate. Note that fixing $\phit$ does not fix $\phi$ because $\phi_{\rm MAD}(\astar)$ evolves with the spin. The accreted rest mass also differs from the net change of the BH mass, $\Delta M_{\rm BH}=\int(e_{\rm HD}+e_{\rm EM})\,dm'$. When the BZ jet efficiency exceeds $100\%$ \citep{tchekhovskoy_efficient_2011}, $e<0$, and the BH loses mass while accreting, as the jet extracts its rotational energy, but its irreducible mass does not decrease \citep{Christodoulou1970}.

Figure~\ref{fig:af-vs-m-phi} shows the final spin as a function of $\phit$ and the ratio of the accreted rest mass to the initial BH mass, $M_{\rm 0, acc}/\mbhzero$, for three initial spins. In the MAD regime, $\phit \gtrsim 1$, the BH needs to accrete $\lesssim 50\%$ of its initial mass to come within $3$--$6\%$ of $\aeq$, for all three initial spins, in agreement with \citet{lowell_rapid_2024}. In the weakly magnetized regime, $\phit \lesssim 0.2$, the BH must accrete several times more: starting from $\abhzero = 0$ at $\phit = 0$, it reaches $\astar \approx 0.95$ after accreting $\mbhzero$ and $\astar \approx 0.99$ after accreting $2\,\mbhzero$. At intermediate flux, $\phit \approx 0.5$, the BH settles at a moderate spin, $\aeq \approx 0.5$, but approaches it at an intermediate rate: starting from $\abhzero = 0$ or $1$, it must accrete $\approx 0.8\,\mbhzero$ to come within $\approx 4\%$ of $\aeq$. Therefore, half of the saturated flux suffices to roughly halve the equilibrium spin, but the accreted mass needed to reach it drops substantially only for $\phit \gtrsim 0.6$.

\section{Numerical implementation of BH evolution} \label{sec:methods}

We implement BH mass and spin evolution in the 3D GRMHD code \texttt{H-AMR} \citep{hamr_paper} by tracking the energy and angular momentum fluxes through the event horizon and periodically updating the Kerr metric. Following previous implementations \citep{janiuk_spinevol_grb_2018,krol_accretion_2021,janiuk_2023,plonka_three-dimensional_2026}, the accretion flow thus evolves through a sequence of stationary Kerr spacetimes. Unlike axisymmetric simulations in full general relativity, which evolve the spacetime self-consistently and find BZ spin-down of collapsar BHs when a strong magnetic field halts the accretion \citep{shibata_2024}, this approach restricts the spacetime to the Kerr family. This restriction is supported by \citet{Kimura2021}, who showed that, in the slow-rotation limit, the time-dependent monopole and dipole components of the BZ backreaction can be represented by evolving the Kerr mass and spin parameters, at rates set by the energy and angular momentum extraction. Therefore, as in \citet{plonka_three-dimensional_2026}, we update the BH mass using the energy flux rather than the rest-mass flux, accounting for the binding energy of the accreted gas and the rotational energy that the jets extract from the BH.

We assume that (i) the work done by the time-dependent metric on the gas is negligible, which we verify a posteriori to be $\lesssim5\%$ of the BH mass growth (Appendix~\ref{app:metric_work}); (ii) the BH mass and spin change little between consecutive metric updates. In our simulations, the BH growth timescale is $t_{\rm gr} = \mbh/\mdot \approx 2\times10^{4}$--$10^{7}\,\rgzero/c$, where $\rgzero=G\mbhzero/c^2$. The rest mass accreted between consecutive metric updates therefore amounts to a fraction $\lesssim 5\times10^{-7}$ of the BH mass [Eq.~\eqref{eq:dt_metric} below]; (iii) the BH spin points along the polar $z$-axis of the grid, aligned or anti-aligned with the orbital angular momentum of the disk, such that the metric remains axisymmetric and we evolve only the $z$-component of the BH angular momentum, $J = J_z$. However, in 3D the accreted angular momentum can have a component perpendicular to the spin axis, even when the disk is initially aligned. Moreover, astrophysical disks are often misaligned with the BH spin: in tidal disruption events (TDEs) and collapsars (Sec.~\ref{sec:discussion}), and in X-ray binaries, where the BH spin axis may be tossed at formation \citep{tauris_tossing_2022}. In MAXI J1820+070, for example, the BH spin is misaligned with the orbit by $\gtrsim 40^\circ$ \citep{poutanen_black_2022}. A misaligned disk supplies less angular momentum along the spin axis and torques the spin direction, which may substantially change the spin evolution. We study the evolution of the spin direction in a follow-up paper.

\texttt{H-AMR} solves the GRMHD equations in conservative form \citep{gammie_harm_2003} on a stationary background spacetime; below we describe how we modify it to allow for metric updates. The spacetime is described by the Kerr metric \citep{kerr_gravitational_1963} in horizon-penetrating Kerr--Schild coordinates, $(t, r, \theta, \varphi)$. The total stress-energy tensor of the gas is given by
\begin{equation}
    T^\mu_\nu = \left( \rho + u_{\rm gas} + p_{\rm gas} + b^2 \right) u^\mu u_\nu + \left( p_{\rm gas} + \frac{b^2}{2} \right) g^\mu_\nu - b^\mu b_\nu,
\end{equation}
where $\rho$ is the rest-mass density, $u_{\rm gas}$ is the internal energy density of the gas, $p_{\rm gas}$ is the gas pressure, $b^\mu$ is the magnetic field 4-vector measured in the comoving frame, and $u^\mu$ is the fluid 4-velocity. Here and in the following sections, we set $G=\mbhzero=c=1$ and absorb the factor of $\sqrt{4\pi}$ into the definition of the magnetic field, such that the magnetic pressure is $b^2/2$.

We split the total stress-energy tensor into hydrodynamic (HD) and electromagnetic (EM) contributions, $T^\mu_\nu = T^{\mu\,({\rm HD})}_{\nu} + T^{\mu\,({\rm EM})}_{\nu}$,
\begin{equation}
    \begin{split}
        T^{\mu\,({\rm HD})}_{\nu} &= \left( \rho + u_{\rm gas} + p_{\rm gas} \right) u^\mu u_\nu + p_{\rm gas} g^\mu_\nu, \\
        T^{\mu\,({\rm EM})}_{\nu} &= b^2 u^\mu u_\nu + \frac{b^2}{2} g^\mu_\nu - b^\mu b_\nu.
    \end{split}
\end{equation}
Conservative GRMHD codes cannot handle vacuum and require numerical floors on the density and internal energy of the gas; \texttt{H-AMR} injects the floor material in the drift frame of the jet \citep{hamr_paper}. The floors are activated mostly in highly magnetized regions, where we cap the magnetization, $\sigma = b^2/\rho$, at $\sigma_{\max} = 150$. The floor material that falls into the BH artificially boosts the HD fluxes through the horizon and can significantly affect the BH evolution \citep{lowell_rapid_2024}. To minimize this contamination, we compute the HD contributions to Eqs.~\eqref{eq:dMdt} and \eqref{eq:dJdt} only over the regions with $\sigma < 2\sigma_{\max}/3$, as in \citet{issa_collapsar_2026} \citep[similar to the cut $\sigma < 30$ for $\sigma_{\max} = 50$ in][]{lowell_rapid_2024}, while the EM contributions are integrated over the entire horizon. We apply this cut to the fluxes that we use to update the metric during the simulation, as well as to all horizon fluxes that we analyze in Sec.~\ref{sec:results}.

The gas accreted by the BH carries energy and angular momentum. The BH mass, $\mbh$, changes through the energy flux across the event horizon,
\begin{equation} \label{eq:dMdt}
    \frac{d\mbh}{dt} = \int T^r_t\, dA_{\theta\varphi},
\end{equation}
and its angular momentum, $J$, changes through the angular momentum flux,
\begin{equation} \label{eq:dJdt}
    \frac{dJ}{dt} = -\int T^r_\varphi\, dA_{\theta\varphi},
\end{equation}
where $dA_{\theta\varphi} = \sqrt{-g}\,d\theta\,d\varphi$ is the area element, with $g$ being the metric determinant. The fluxes are defined to be positive into the BH: for inflowing prograde gas, $T^r_t > 0$ and $T^r_\varphi < 0$. We evaluate the integrals in the radial grid cell right outside the instantaneous outer horizon, $r_{\rm H} = m\left(1+\sqrt{1-\astar^2}\right)$, where $m \equiv \mbh/\mbhzero$.

The dimensionless spin of the BH, $\astar = J/\mbh^2$, evolves as
\begin{equation} \label{eq:dadt}
    \frac{d\astar}{dt} = \frac{1}{\mbh^2}\frac{dJ}{dt} - \frac{2\astar}{\mbh}\frac{d\mbh}{dt}.
\end{equation}
Accretion of angular momentum spins the BH up, while accretion of energy spins it down. It is instructive to divide by the rest-mass accretion rate, $\mdot = -\int \rho u^r dA_{\theta\varphi}$, and define the specific angular momentum and energy of the accreted gas,
\begin{equation} \label{eq:l_and_e_def}
    l = \frac{1}{\mbh \mdot} \frac{dJ}{dt}, \quad e = \frac{1}{\mdot} \frac{d\mbh}{dt},
\end{equation}
such that Eq.~\eqref{eq:dadt} becomes $d\astar/dt = (\mdot/\mbh)(l - 2\astar e)$ \citep{bardeen_kerr_1970,gammie_black_2004,shapiro_spin_2005}. Computing Eqs.~\eqref{eq:dMdt} and \eqref{eq:dJdt} with $T^{\mu\,({\rm HD})}_{\nu}$ or $T^{\mu\,({\rm EM})}_{\nu}$ instead of $T^\mu_\nu$ gives the HD and EM contributions, $l_{\rm HD}$, $e_{\rm HD}$, $l_{\rm EM}$, and $e_{\rm EM}$.

We keep $\mbhzero = 1$ in code units, so that the metric depends explicitly on the dimensionless BH mass, $m$, and on the spin parameter, $a \equiv \astar m$ (in units of $\rgzero$). After each metric update step, we recompute the metric tensor, $g_{\mu\nu}$,
\begin{equation}
    \begin{split}
        g_{tt} &= -\left(1 - \frac{2mr}{\Sigma}\right), \\
        g_{tr} &= g_{rt} = \frac{2mr}{\Sigma}, \\
        g_{t\varphi} &= g_{\varphi t} = -\frac{2mra \sin^2\theta}{\Sigma}, \\
        g_{rr} &= 1 + \frac{2mr}{\Sigma}, \\
        g_{r\varphi} &= g_{\varphi r} = -a \sin^2\theta \left(1 + \frac{2mr}{\Sigma}\right), \\
        g_{\theta\theta} &= \Sigma, \\
        g_{\varphi\varphi} &= \sin^2\theta \left( r^2 + a^2 + \frac{2mra^2 \sin^2\theta}{\Sigma} \right),
    \end{split}
\end{equation}
where $\Sigma = r^2 + a^2 \cos^2\theta$. The coordinates remain unchanged between updates, while the location of the event horizon changes. After the metric update, we recompute the connection coefficients, $\Gamma^\mu_{\nu\lambda}$, and the metric determinant, $g$.

Across a metric update, we hold the conserved variables, $\mathcal{U} = \left[ \sqrt{-g}\rho u^t, \sqrt{-g} T^t_t, \sqrt{-g} T^t_i, \sqrt{-g} B^i \right]$, fixed, where $B^i$ is the magnetic field 3-vector in the coordinate frame. We then recover the primitive variables, $\mathcal{P}$, from $\mathcal{U}$ using the updated metric \citep{noble_primitive_2006}. In Kerr--Schild coordinates, $\sqrt{-g} = \Sigma \sin\theta$ depends on $a$ but not on $m$. Holding $\mathcal{U}$ fixed exactly conserves the rest mass, the magnetic flux (so that the constrained transport scheme keeps $\nabla\cdot B = 0$), and the angular momentum of the gas: the continuity and induction equations contain no source terms, and the source term in the angular momentum equation, $\frac{1}{2}\sqrt{-g}\,T^{\alpha\lambda}\partial_\varphi g_{\alpha\lambda}$, vanishes because the metric remains axisymmetric. The only term this procedure omits is the source term in the energy equation, $\frac{1}{2}\sqrt{-g}\,T^{\alpha\lambda}\partial_t g_{\alpha\lambda}$, which describes the work done on the gas by the time-dependent metric. The total energy and angular momentum of the gas and the BH, including the outer-boundary fluxes, are conserved to $\lesssim1\%$ in most simulations, and to within $\lesssim1.7\%$ (energy) and $\lesssim2.7\%$ (angular momentum) in all of them (Table~\ref{tab:simulation_setup}). In Appendix~\ref{app:metric_work}, we show that the energy omitted by neglecting the work done by the metric is $\lesssim 4\%$ of the BH mass growth, or $\lesssim 0.02\,\mbhzero c^2$. This is smaller than the non-conservation of total energy in the MAD and tMAD models, but larger than that in the SANE models, which otherwise conserve energy to $\lesssim 0.07\%$.

We update the metric at intervals of
\begin{equation} \label{eq:dt_metric}
    \Delta t_{\rm metric} = \max\left(10^{-2}\rgzero/c,\, \Delta t \right),
\end{equation}
where $\Delta t$ is the fluid timestep on the grid. We integrate $\mbh$ and $\astar$ in time with a second-order Adams--Bashforth scheme, which uses their time derivatives at the current and previous updates, $t_n$ and $t_{n-1}$, to obtain their values at $t_{n+1}$:
\begin{equation}
    \begin{split}
        \mbh^{n+1} &= \mbh^n + \Delta t_{\rm metric, curr} \left( w_{\rm curr}\, \dot{M}_{\rm BH}^n - w_{\rm prev}\, \dot{M}_{\rm BH}^{n-1} \right), \\
        \astar^{n+1} &= \astar^n + \Delta t_{\rm metric, curr} \left( w_{\rm curr}\, \dot{a}_\star^n - w_{\rm prev}\, \dot{a}_\star^{n-1} \right),
    \end{split}
\end{equation}
where $\dot{M}_{\rm BH}^n$ and $\dot{a}_\star^n$ are given by Eqs.~\eqref{eq:dMdt} and \eqref{eq:dadt} at $t_n$, $\Delta t_{\rm metric, curr} = t_{n+1} - t_n$, and $\Delta t_{\rm metric, prev} = t_n - t_{n-1}$. The weights, $w_{\rm curr} = 1 + \Delta t_{\rm metric, curr}/(2 \Delta t_{\rm metric, prev})$ and $w_{\rm prev} = \Delta t_{\rm metric, curr}/(2 \Delta t_{\rm metric, prev})$, account for the variation of the update interval between updates, and reduce to $3/2$ and $1/2$ for a constant interval.

As the BH evolves, both its outer and inner Kerr horizons, $r_\pm = m\left(1 \pm \sqrt{1-\astar^2}\right)$, move with respect to the fixed grid. At fixed spin magnitude, both horizon radii scale with \(m\). At fixed mass, the outer horizon shrinks and the inner horizon grows as $\lvert\astar\rvert$ increases. For the inner radial boundary to remain causally disconnected from the simulation domain, it must lie between the two horizons. The inner edge of our grid is fixed at $0.8\,r_{\rm H}(\abhzero)$ (Sec.~\ref{sec:numerical_setup}), and we additionally define a moving inner radius, $r_{\rm b} = 0.7\,r_{\rm H}$. We overwrite the gas at $r < r_{\rm b}$ with floor values and set the fluxes between these cells and their neighbors to zero. Initially, $r_{\rm b}$ lies inside the inner edge of the grid, and it starts removing grid cells only once $r_{\rm H}$ has grown by more than a factor of $0.8/0.7 \approx 1.14$. For all spins reached in our simulations, $r_{\rm b}$ lies outside the inner horizon\footnote{The ratio $r_-/r_+ = \bigl(1-\sqrt{1-\astar^2}\bigr)/\bigl(1+\sqrt{1-\astar^2}\bigr)$ increases with $\lvert\astar\rvert$, and the inner horizon reaches $r_{\rm b} = 0.7\,r_{\rm H}$ at $\lvert\astar\rvert \approx 0.984$. Simulations that approach maximal spin would therefore require a larger $r_{\rm b}/r_{\rm H}$, or an $r_{\rm b}$ placed between $r_-$ and $r_+$, at the cost of fewer cells between $r_{\rm b}$ and $r_{\rm H}$.}, and with our logarithmic radial grid ($\Delta \ln r \approx 0.04$) there are $\approx 9$ cells between $r_{\rm b}$ and $r_{\rm H}$.

\section{Simulations}\label{sec:simulations}
\subsection{Numerical setup} \label{sec:numerical_setup}

\begin{table*}[htbp]
    \centering
    \caption{Summary of the simulation setup parameters. The table lists the initial magnetic field geometry [Eq.~\eqref{eq:vector_potential}], the ratio of the maximum gas pressure to the maximum magnetic pressure $\beta_{\rm max}$, the initial torus-to-BH mass ratio $M_{\rm torus}/\mbhzero$, the initial BH spin $\abhzero$, the total simulation runtime in units of $10^4\,\rgzero/c$, the accretion state of the disk: MAD, transient MAD (tMAD), or SANE, and the largest relative deviations of the total energy, $(\Delta E/E_0)_{\max}$, and angular momentum, $(\Delta L/L_0)_{\max}$, over the simulation runtime, where $E_0$ and $L_0$ are the initial total energy (including rest mass) and angular momentum of the gas and the BH. Simulations with fixed BH mass and spin are marked with $\dagger$.}
    \label{tab:simulation_setup}
    \begin{tabular*}{\textwidth}{@{\extracolsep{\fill}}cccccccc}
        \hline\hline
        Field geometry & $\beta_{\rm max}$ & $M_{\rm torus}/\mbhzero$ & $\abhzero$ & Runtime [$10^4\,\rgzero/c$] & Accretion state & $(\Delta E/E_0)_{\max}$ [\%] & $(\Delta L/L_0)_{\max}$ [\%] \\
        \hline
        Small loop & $100$ & $4.31$ & $0.1$ & $4.07$ & SANE & $0.01$ & $0.02$ \\
        Small loop & $100$ & $4.95$ & $0.5$ & $3.99$ & SANE & $0.02$ & $0.06$ \\
        Small loop & $100$ & $0.43$ & $0.1$ & $4.12$ & SANE & $0.01$ & $0.02$ \\
        Small loop & $10$ & $0.46$ & $0.3$ & $4.06$ & SANE & $0.07$ & $0.23$ \\ \hline
        Large loop & $10$ & $3.88$ & $-0.2$ & $5.36$ & MAD & $0.59$ & $0.14$ \\
        Large loop & $10$ & $4.31$ & $0.1$ & $4.70$ & MAD & $0.55$ & $0.11$ \\
        Large loop & $10$ & $4.95$ & $0.5$ & $4.52$ & MAD & $0.66$ & $0.35$ \\
        Large loop & $10$ & $0.40$ & $-0.1$ & $9.82$ & MAD & $0.81$ & $0.41$ \\
        Large loop & $10$ & $0.35$ & $-0.5$ & $9.19$ & MAD & $0.99$ & $1.35$ \\
        Large loop & $10$ & $0.49$ & $0.5$ & $1.94$ & MAD & $0.91$ & $0.91$ \\
        Large loop & $10$ & $0.56$ & $0.9$ & $3.38$ & MAD & $1.66$ & $2.73$ \\ \hline
        Large loop & $100$ & $4.31$ & $0.1$ & $4.60$ & tMAD & $0.51$ & $0.01$ \\
        Large loop & $100$ & $4.95$ & $0.5$ & $4.48$ & tMAD & $0.43$ & $0.12$ \\ \hline
        Medium loop & $100$ & $4.31$ & $0.1$ & $8.69$ & tMAD & $0.57$ & $0.10$ \\
        Medium loop & $100$ & $4.54$ & $0.25$ & $5.28$ & tMAD & $0.38$ & $0.12$ \\
        Medium loop & $100$ & $4.95$ & $0.5$ & $8.47$ & tMAD & $0.57$ & $0.12$ \\
        Medium loop & $100$ & $5.72$ & $0.94$ & $8.17$ & tMAD & $0.52$ & $0.32$ \\ \hline
        Large loop & $100$ & $4.95$ & $0.5$ & $3.84$ & MAD$^{\dagger}$ & $0.89$ & $0.15$ \\ \hline
        Medium loop & $100$ & $4.31$ & $0.1$ & $3.14$ & MAD$^{\dagger}$ & $0.47$ & $0.01$ \\
        Medium loop & $100$ & $4.54$ & $0.25$ & $2.02$ & MAD$^{\dagger}$ & $0.36$ & $0.01$ \\
        Medium loop & $100$ & $4.95$ & $0.5$ & $2.98$ & MAD$^{\dagger}$ & $0.57$ & $0.28$ \\
        Medium loop & $100$ & $5.72$ & $0.94$ & $1.61$ & MAD$^{\dagger}$ & $0.39$ & $0.49$ \\ \hline
    \end{tabular*}
\end{table*}

We perform a suite of simulations to compute the spin evolution in accreting BH systems. The simulations are initialized with a BH of initial mass $\mbhzero$ and spin $\abhzero$ surrounded by a Fishbone--Moncrief torus \citep{fishbone_relativistic_1976} with the inner edge at $r_{\rm in} = 20\,\rgzero$ and the pressure maximum at $r_{\rm max} = 41\,\rgzero$. We set $\rho_{\max} = 1$ in code units and choose $\rho_{\rm scale}$ to set $M_{\rm torus}/\mbhzero$. Because we neglect self-gravity and cooling, the GRMHD equations are invariant under a rescaling of the gas density, such that at a fixed metric the gas evolves identically for any torus mass. The torus mass enters only when we convert the horizon fluxes into changes of the BH mass and spin, through the physical density scale, $\rho_{\rm scale}$, which sets $M_{\rm torus}/\mbhzero$ (Table~\ref{tab:simulation_setup}). Thus, at fixed BH mass and spin, $\rho_{\rm scale}$ changes the normalization of the accretion rate without changing the flow structure. When BH evolution is enabled, it also sets the timescale on which the metric changes, allowing the subsequent flow evolution to differ. Because we neglect gas self-gravity, the massive tori serve as idealized reservoirs that accelerate BH evolution rather than self-consistent models of self-gravitating disks. For gas thermodynamics, we assume an ideal gas equation of state, $p_{\rm gas} = (\gamma_{\rm ad} - 1) u_{\rm gas}$, with the adiabatic index $\gamma_{\rm ad} = 4/3$. We add small random perturbations to the internal energy of the gas to seed the growth of the magnetorotational instability \citep[MRI;][]{mri_paper} that drives the accretion.

We thread the torus with a poloidal magnetic field, which we set through the azimuthal component of the vector potential, $A_\varphi$. To study the effect of the magnetic field topology, we consider three geometries: a small loop confined to the densest part of the torus, a medium loop, and a large loop that extends over a larger portion of the torus:
\begin{equation} \label{eq:vector_potential}
    \begin{split}
        A_{\varphi; \rm small} &= \max\left[ \frac{\rho}{\rho_{\max}} - 0.2, 0\right], \\
        A_{\varphi; \rm med} &= \max\left( \frac{\rho}{\rho_{\max}} - 5\times 10^{-4}, 0 \right) r^5, \\
        A_{\varphi; \rm large} &= \max\left[ \frac{\rho}{\rho_{\max}} \left( \frac{r\sin\theta}{r_{\rm in}}\right)^3 e^{-r/r_{\rm t}} - 0.2, 0\right], 
    \end{split}
\end{equation}
where $r_{\rm t} = 400\,\rgzero$. We vary the magnetic field strength through the ratio of the maximum gas pressure to the maximum magnetic pressure in the torus, $\beta_{\rm max} \equiv \max p_{\rm gas} / \max p_{\rm mag}$. Table~\ref{tab:simulation_setup} summarizes the simulation parameters.

We use logarithmic Kerr--Schild coordinates, $\{X^0, X^1, X^2, X^3\} = \{t, \ln r, \theta, \varphi\}$, uniformly spaced in $X^1$, $X^2$, and $X^3$. The 3D grid extends from $r = 0.8\,r_{\rm H}(\abhzero)$ to $10^5\,\rgzero$ in the radial direction and covers $\theta \in [0, \pi]$ and the whole $\varphi \in [0, 2\pi)$, with a resolution of $N_r \times N_\theta \times N_\varphi = 288 \times 192 \times 64$. We use outflow boundary conditions at the inner and outer radial boundaries, transmissive boundary conditions in $\theta$, and periodic boundary conditions in $\varphi$.

\subsection{Results} \label{sec:results}

\subsubsection{Time evolution of the BHs}

\begin{figure}
    \centering
    \includegraphics[width=\linewidth]{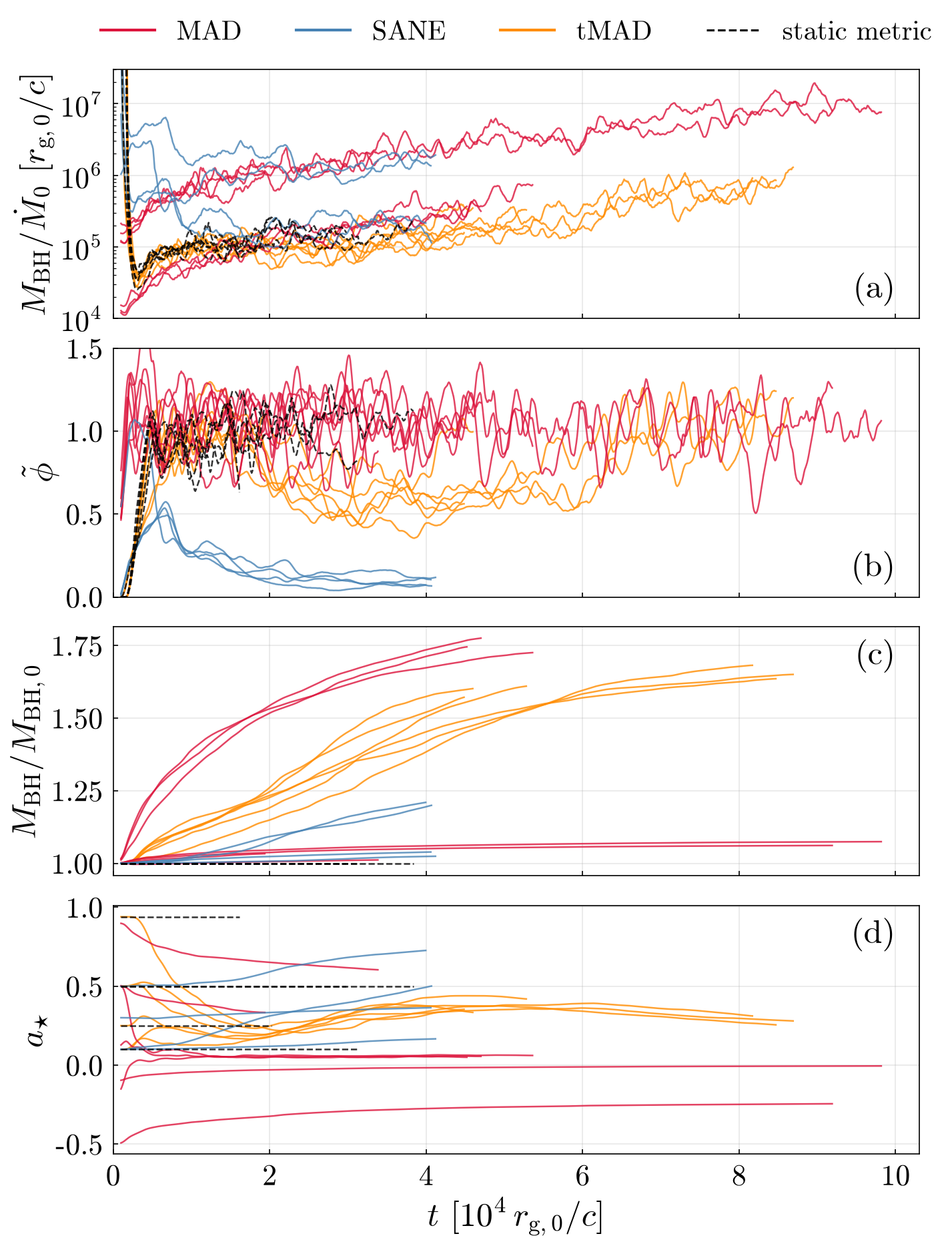}
    \caption{Time evolution of the quantities on the BH horizon in all simulations. {\bf (a)}: the BH mass growth timescale. {\bf (b)}: MAD parameter. {\bf (c)}: mass-energy of the BH. {\bf (d)}: spin evolution toward equilibrium spin. Black dashed lines represent static metric runs.}
    \label{fig:time-evolution-plot}
\end{figure}

Figure~\ref{fig:time-evolution-plot}(a) depicts the time evolution of the BH mass growth timescale, defined as the instantaneous BH mass, $\mbh$, divided by the rest-mass accretion rate, $\mdot$, measured at the event horizon,
\begin{equation}
    \mdot = - \int \rho  u^r dA_{\theta\varphi} \Big|_{\sigma < 2\sigma_{\max}/3}.
\end{equation}

The two torus-mass scales $M_{\rm torus}/\mbhzero$ produce two separate tracks of increasing $\mbh/\mdot$. The corresponding BH growth times span $2\times10^4$--$10^7\,\rgzero/c$. Fig.~\ref{fig:time-evolution-plot}(b) shows the MAD parameter, $\phit(\astar)$, where
\begin{equation} \label{eq:phi}
    \phi = \frac{\sqrt{4\pi}\,\Phi_{\rm H}^{\rm HL}}{\sqrt{\mdot\, \rg^2 c}} = \frac{\sqrt{4\pi}}{2 m \sqrt{\mdot}} \int |B^r|\, dA_{\theta\varphi},
\end{equation}
where $\Phi_{\rm H}^{\rm HL}=\frac{1}{2}\int|B^r|\,dA_{\theta\varphi}$ is the horizon magnetic flux in the Heaviside--Lorentz units of the code (Sec.~\ref{sec:methods}), such that $\Phi_{\rm H}=\sqrt{4\pi}\,\Phi_{\rm H}^{\rm HL}$ in Eq.~\eqref{eq:phi0}, and the last equality uses $G=c=\mbhzero=1$, with $\rg=m$. Omitting $m$ would overestimate $\phi$ by a factor of $m$, which reaches $\approx 1.75$ in our most rapidly accreting MADs [Fig.~\ref{fig:time-evolution-plot}(c)].

Large-loop models with stronger initial magnetic fields, $\beta_{\max}=10$, reach the MAD state within the first few thousand $\rgzero/c$ and remain there until the end of the simulation, owing to their large total flux reservoir. Small-loop models, regardless of $\beta_{\max}$, never saturate the horizon flux. The large-loop models with weaker initial fields, $\beta_{\max}=100$, are already near a MAD state at $t \lesssim 10^4\,\rgzero/c$, but leave the MAD state at $t \approx 1$--$2\times10^4\,\rgzero/c$ (transient MADs, tMADs). One model returns to $\phi \approx \phi_{\rm MAD}$ late in the simulation owing to the drop in $\mdot$. \citet{wong_sane_2026} demonstrated that relying on $\phi/\phi_{\rm MAD}$ as the sole diagnostic of the MAD state can be misleading: a flow can reach MAD-level horizon flux and jet power without exhibiting the flux eruptions characteristic of MADs \citep{igumenshchev_magnetically_2008,tchekhovskoy_efficient_2011,jacquemin-ide_demystifying_2025}. In Appendix~\ref{app:mad_state}, we compare the radial profiles of the enclosed magnetic flux to verify that disks that reach the MAD state exhibit flux eruptions, while those that leave it stop erupting.

Figure~\ref{fig:time-evolution-plot}(c),(d) show the evolution of the BH masses and spins. The high-$\mdot$ MAD models with $\abhzero = 0.1$ and $0.5$ both spin down and level off at an equilibrium spin $\aeq \approx 0.05$--$0.06$, between the estimates for radiatively inefficient MADs by \citet{narayan_jets_2022} and \citet{lowell_rapid_2024}, $0.035$ and $0.07$, respectively. In low-$\mdot$ MAD models, the BHs spin up or down toward $\aeq$, but they gain only $\approx 7\%$ of their initial mass, too little to reach $\aeq$, so that their final spin is set by the available mass reservoir. BHs in tMAD models spin down while $\phi \approx \phi_{\rm MAD}$ and spin up once the disk leaves the MAD state. BHs in all SANE models spin up.

Comparing the evolving-BH simulations with their fixed-metric counterparts, we find similar evolution of the absolute horizon magnetic flux and the maximum enclosed disk flux, but systematically higher mass accretion rates in the evolving runs. Since $\phi \propto \Phi_{\rm H}/(m \dot M_0^{1/2})$ [Eq.~\eqref{eq:phi}], both BH growth and the enhanced accretion rate reduce the normalized horizon flux. Thus, the transition out of the MAD state does not require the evolving-BH simulations to lose more absolute magnetic flux. Instead, the available flux becomes insufficient to sustain magnetic arrest as the BH grows and accretes more rapidly.

\subsubsection{Black hole spin evolution}

\begin{table*}[htbp]
    \centering
    \caption{Fitting functions used in Sec.~\ref{sec:results}. The HD fits are normalized by the NT specific energy and angular momentum, $e_{\rm NT}(\astar)$ and $l_{\rm NT}(\astar)$ (Appendix~\ref{app:nt}). The spin-up parameter is $s = l_{\rm HD} + l_{\rm EM} - 2\astar (e_{\rm HD} + e_{\rm EM})$, and $\aeq$ is the root of $s(\astar, \phit) = 0$; its fit is valid for $0 \le \phit \lesssim 1.2$.}
    \label{tab:fits}
    \begin{tabular}{llll}
        \hline\hline
        Quantity & Fitting function & Coefficients & Figure \\
        \hline
        $\phi_{\rm MAD}$ & $p_0 + p_1 \astar + p_2 \astar^2 + p_3 \astar^3$ & $p_0 = 52.6$, $p_1 = 34$, $p_2 = -14.9$, $p_3 = -20.2$ \citep{narayan_jets_2022} & \ref{fig:phi-mad-fit} \\
        $\aeq$ & $\sum_{n=0}^{4} q_n \phit^n$ & $q_0 = 1$, $q_1 = -0.13$, $q_2 = -3.55$, $q_3 = 4.27$, $q_4 = -1.47$ & \ref{fig:aeq-fit} \\
        $e_{\rm EM}$ & $-\phit^2 \left( c_2 \astar^2 + c_4 \astar^4 \right)$ & $c_2 = 0.12$, $c_4 = 2.58$ & \ref{fig:specific-fluxes-fit}(a) \\
        $l_{\rm EM}$ & $-\phit^2 \left( d_1 \astar + d_3 \astar^3 \right)$ & $d_1 = 8.23$, $d_3 = 5.02$ & \ref{fig:specific-fluxes-fit}(b) \\
        $e_{\rm HD}$ & $e_{\rm NT}(\astar) \left( h_0 + h_1 \phit \right)$ & $h_0 = 1.07$, $h_1 = -0.05$ & \ref{fig:specific-fluxes-fit}(c) \\
        $l_{\rm HD}$ & $l_{\rm NT}(\astar) \left( g_0 + g_1 \phit \right)$ & $g_0 = 1$, $g_1 = -0.63$ & \ref{fig:specific-fluxes-fit}(d) \\
        \hline
    \end{tabular}
\end{table*}

\begin{figure*}
    \centering
    \includegraphics[width=\linewidth]{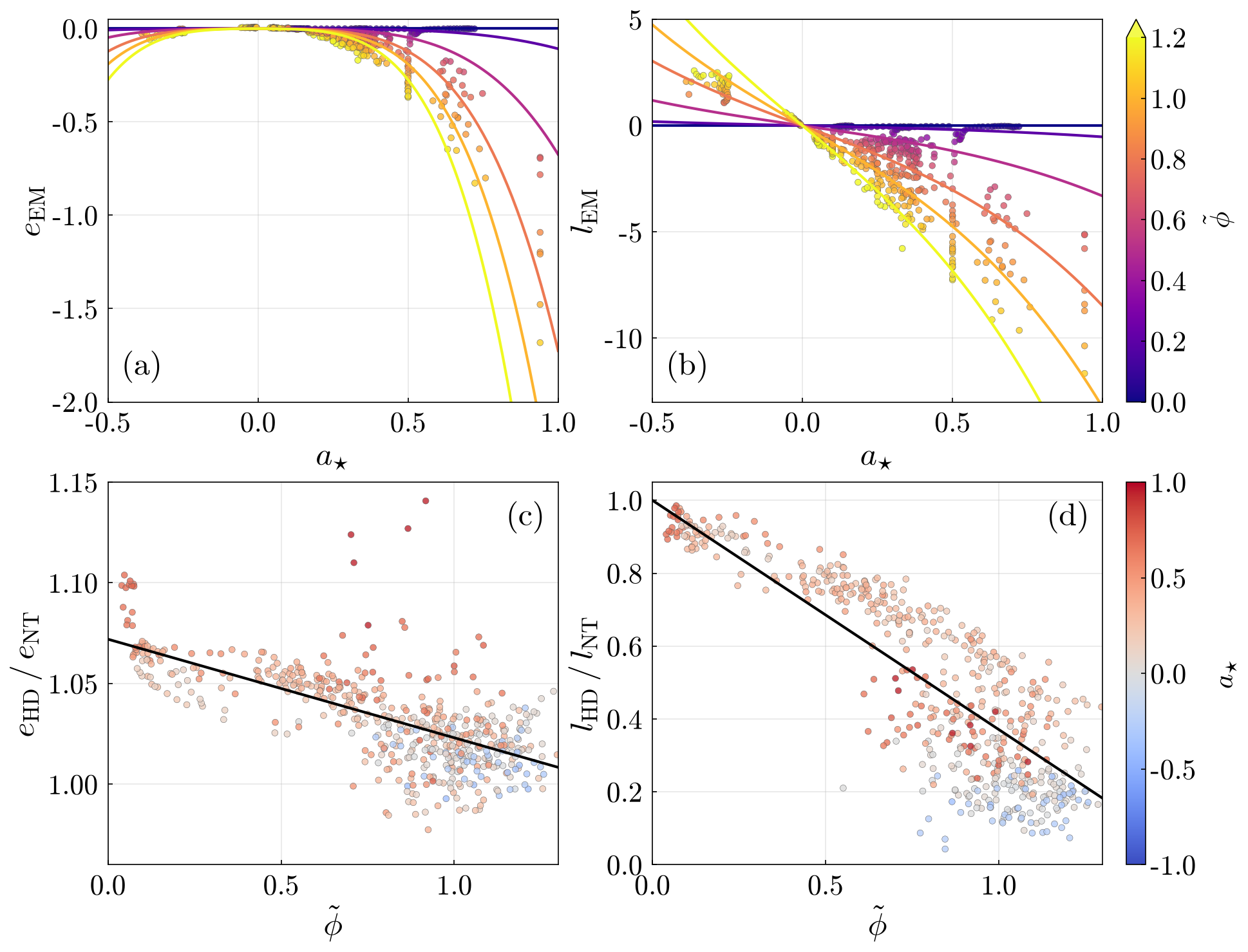}
    \caption{Specific energy and angular momentum fluxes. Panels (a,b) show the EM fluxes as functions of spin, $\astar$, colored by the MAD parameter, $\phit$. Panels (c,d) show the HD fluxes, normalized by their NT values, as functions of $\phit$, colored by $\astar$. Curves show the fits in Table~\ref{tab:fits}.}
    \label{fig:specific-fluxes-fit}
\end{figure*}

We use our simulations to construct a prescription for BH spin evolution across the range of horizon magnetic fluxes, $\phit \lesssim 1.2$, connecting the SANE and MAD limits. We reduce each simulation to a set of approximately independent measurements of the specific energy and angular momentum fluxes, averaged over time windows and binned in $(\astar, \phit)$ (Fig.~\ref{fig:specific-fluxes-fit}; Appendix~\ref{app:fits}). Using Eq.~\eqref{eq:dadt}, we express $s$ through the HD and EM contributions to the specific fluxes defined in Eq.~\eqref{eq:l_and_e_def},
\begin{equation}
    s = \sum_{\rm X = \{HD, EM\}} ( l_{\rm X} - 2 \astar e_{\rm X} ).
\end{equation}

The BZ jet efficiency depends on the horizon magnetic flux and the BH spin as \citep{blandford_electromagnetic_1977,tchekhovskoy_2010,tchekhovskoy_agn_2015}
\begin{equation}
    \eta_{\rm BZ} = \frac{\kappa}{4\pi}\, \omega_{\rm H}^2\, \phi^2 f(\omega_{\rm H}),
\end{equation}
where $\kappa$ is a constant that depends on the magnetic field geometry ($\kappa \approx 0.05$ for a split monopole), $\phi$ is given by Eq.~\eqref{eq:phi}, $\omega_{\rm H} \equiv \Omega_{\rm H} \rg / c = \astar / \left[ 2 \left( 1 + \sqrt{1 - \astar^2} \right) \right]$ is the dimensionless angular frequency of the horizon, and $f(\omega_{\rm H}) = 1 + 1.38\, \omega_{\rm H}^2 - 9.2\, \omega_{\rm H}^4$ is the high-spin correction \citep{tchekhovskoy_2010}. This guides our choice of the functional form for the fit of $e_{\rm EM}$ [Fig.~\ref{fig:specific-fluxes-fit}(a)],
\begin{equation}
    e_{\rm EM; fit} (\astar, \phit) = -\phit^2 \left( 0.12 \astar^2 + 2.58 \astar^4 \right).
\end{equation}

Following \citet{moderski_black_1996} and \citet{lowell_rapid_2024}, the EM energy and angular momentum fluxes are related through the angular frequency of the magnetic field lines, $\Omega_{\rm F} = k \Omega_{\rm H}$, with $k = 1/2$ for the BZ monopole; in our dimensionless units, $l_{\rm EM} = e_{\rm EM}/(k\,\omega_{\rm H})$. \citet{lowell_rapid_2024,lowell_evidence_2025} find $k \sim 0.1$--$0.5$ for a range of hot and cold MADs, with a weak spin dependence. Because $e_{\rm EM; fit} \propto c_2\,\astar^2 + c_4\,\astar^4$ and $\omega_{\rm H} \approx \astar/4$ at low $\astar$, $e_{\rm EM; fit}/\omega_{\rm H}$ contains terms $\propto \astar$ and $\astar^3$, which motivates our choice [Fig.~\ref{fig:specific-fluxes-fit}(b)],
\begin{equation}
    l_{\rm EM; fit} (\astar, \phit) = -\phit^2 \left( 8.23\,\astar + 5.02\,\astar^3 \right).
\end{equation}
Together, these fits imply $k = e_{\rm EM; fit}/(\omega_{\rm H}\, l_{\rm EM; fit}) \approx 0.07$, $0.3$, and $0.5$ at $\astar = 0.1$, $0.5$, and $0.9$, respectively, a stronger spin dependence than that reported by \citet{lowell_rapid_2024}. At small spins, the electromagnetic energy term contributes little to $s$, while the electromagnetic angular momentum flux remains important.

We express the HD contributions to the specific fluxes relative to those of an NT disk, $e_{\rm NT}(\astar)$ and $l_{\rm NT}(\astar)$, i.e., the specific energy and angular momentum of circular geodesics at the innermost stable circular orbit \citep[ISCO;][]{bardeen_rotating_1972}, which we provide in Appendix~\ref{app:nt}. We find that $e_{\rm HD}/e_{\rm NT}$ is well approximated by [Fig.~\ref{fig:specific-fluxes-fit}(c)]
\begin{equation}
    e_{\rm HD; fit} (\astar, \phit) = e_{\rm NT} (\astar) \left( 1.07 - 0.05\,\phit \right).
\end{equation}
The specific angular momentum delivered by the accretion flow decreases as it becomes more magnetized, and we choose [Fig.~\ref{fig:specific-fluxes-fit}(d)]
\begin{equation}
    l_{\rm HD; fit} (\astar, \phit) = l_{\rm NT}(\astar) \left( 1 - 0.63\,\phit \right).
\end{equation}
At $\phit = 0$, these fits give $e_{\rm HD; fit} = 1.07\, e_{\rm NT}$ and $l_{\rm HD; fit} = l_{\rm NT}$, and therefore approximately recover the NT limit.

\begin{figure}
    \centering
    \includegraphics[width=\linewidth]{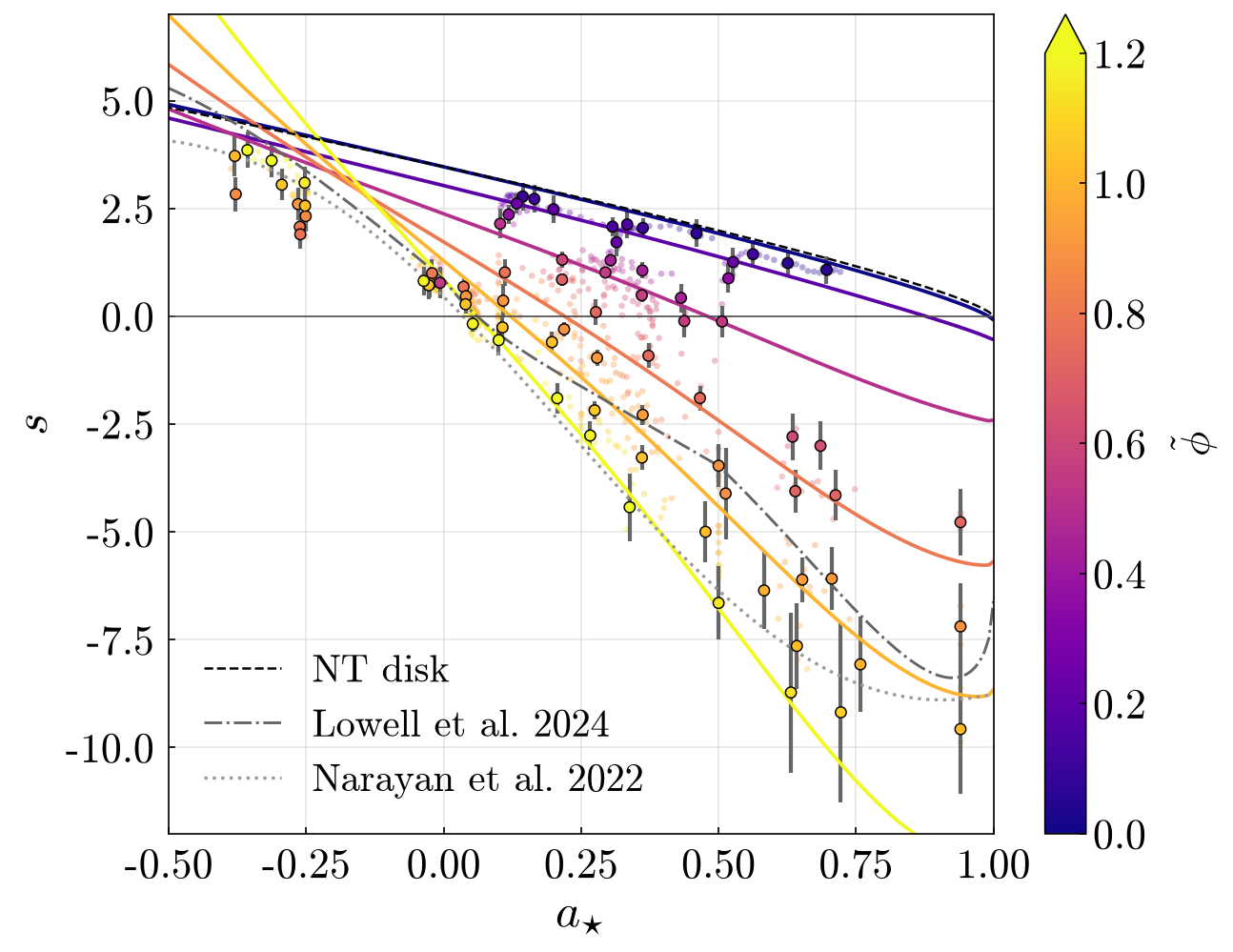}
    \caption{Spin-up parameter as a function of spin, with points colored by the MAD parameter, $\phit$. Solid curves show the fit [Eq.~\eqref{eq:spinup_fit}, Table~\ref{tab:fits}] at fixed $\phit$; the dashed curve shows the NT disk. The fits from \citet{narayan_jets_2022,lowell_rapid_2024} are shown for comparison.}
    \label{fig:spinup-fit}
\end{figure}

Finally, we combine the components into a single prescription for the spin-up parameter,
\begin{equation} \label{eq:spinup_fit}
    s = s_{\rm HD} + s_{\rm EM} = l_{\rm HD} + l_{\rm EM} - 2\astar (e_{\rm HD} + e_{\rm EM}),
\end{equation}
shown in Figure~\ref{fig:spinup-fit}; Table~\ref{tab:fits} summarizes the fitting functions. Our fit approximately recovers the NT limit (dashed line) and largely agrees with \citet{narayan_jets_2022} and \citet{lowell_rapid_2024} for hot MADs at moderate spins. The fitted measurements sample $-0.4 \lesssim \astar \lesssim 0.94$, but not uniformly: retrograde spins are sampled only at $\phit \gtrsim 0.75$, and weakly magnetized flows, $\phit \lesssim 0.3$, only at $0.1 \lesssim \astar \lesssim 0.7$. Outside these ranges, including $\aeq$ at $\phit \lesssim 0.3$, the fits are extrapolations.

\section{Conclusions and Implications} \label{sec:discussion}

In this paper, we implemented BH mass and spin evolution in the GRMHD code \texttt{H-AMR} and studied how BHs evolve in magnetized accretion flows by simulating idealized tori, which represent hot, radiatively inefficient flows. Our fits for the spin-up parameter, $s(\astar,\phit)$, and the equilibrium spin, $\aeq(\phit)$, can be used as prescriptions in galaxy-scale and cosmological simulations.

The equilibrium spin, $\aeq$, is a continuous, steep function of the MAD parameter, $\phit$ (see Fig.~\ref{fig:aeq-fit}). In the weakly magnetized limit, $\phit < 0.2$, our fits show $\aeq \gtrsim 0.88$, in agreement with steady-state models of thin disks \citep{bardeen_kerr_1970,thorne_disk-accretion_1974}, hot advection-dominated flows \citep{Popham1998,Benson2009}, and with GRMHD simulations of SANE disks \citep{gammie_black_2004,chatterjee_mad_eruptions_2022,dhruv_survey_2025}.
At intermediate $\phit$, $\aeq$ decreases monotonically from $\aeq\approx 0.88$ at $\phit = 0.2$ to $\aeq\approx 0.5$ at $\phit = 0.5$ and $\approx 0.21$ at $\phit = 0.8$.
As we approach the MAD limit, $\phit \gtrsim 1$, we find $\aeq \approx 0.06$, consistent with the results obtained from simulations of hot MADs in a stationary metric \citep{narayan_jets_2022,lowell_rapid_2024}.

Whether a BH actually reaches $\aeq$ depends on how much rest mass it accretes, $M_{0,\rm acc}$, compared with its own mass (see Fig.~\ref{fig:af-vs-m-phi}). 
Starting from $\abhzero=0$, a weakly magnetized flow must supply $M_{0,\rm acc} \gtrsim \mbhzero$ to approach $\aeq$. MADs spin down their BHs more efficiently, requiring only $M_{0,\rm acc} \lesssim 0.5 \mbhzero$ to reach their low $\aeq$.
The required mass drops substantially only for $\phit \gtrsim 0.6$, as a flow at intermediate $\phit$ changes the BH spin almost as slowly as a SANE, but toward a much lower spin.

Accounting for the BH growth also changes the magnetic state of the disk. The flux needed for a MAD, $\Phi_{\rm H} \approx \phi_{\rm MAD} (\dot M_0 c)^{1/2} r_{\rm g} \propto M_{\rm BH} \dot M_0^{1/2}$ \citep[see also][]{tchekhovskoy_magnetic_2015}, grows with the BH mass. Therefore, disks with a finite flux reservoir that remain MAD around a fixed BH leave the MAD state once the BH has grown by $\approx 10$--$25\%$, after which the BH spins up toward a higher $\aeq$. In contrast, disks with a larger flux reservoir remain MAD even as their BHs grow by $\approx 75\%$. A growing BH thus remains MAD only if the flux it accumulates keeps pace with $M_{\rm BH} \dot M_0^{1/2}$.

\subsection{Implications for spins in BH-powered engines} \label{sec:disc_sources}

\subsubsection{Collapsars}

In collapsars, accretion can substantially change the BH spin within the engine lifetime. The BH forms when the hyperaccreting proto-neutron star (PNS) collapses, with an initial mass $\mbhzero \approx 3\,\msun$ \citep{oconnor_black_2011} and a natal spin $\abhzero \lesssim 0.5$, even if the PNS rotates rapidly \citep{Gottlieb2024}. Adopting the upper end of typical GRB engine powers and jet efficiencies limits the accretion rate to $\mdot \lesssim 0.4\,\msun\,{\rm s}^{-1}$ \citep{Gottlieb2026}, and the rest mass accreted after jet launch to $M_{0,\rm acc} \lesssim \mbhzero$ \citep{gottlieb_collapsar_2023}. For the jet to emerge from the ergosphere, the horizon must hold a dynamically important magnetic flux, $\phi \gtrsim 10$--$20$ \citep{issa_magnetically_2025}, or $\phit \gtrsim 0.2$--$0.3$ for $\phi_{\rm MAD} \approx 60$ at moderate spins. Whether this flux is inherited from the PNS \citep{Gottlieb2024} or generated by a disk dynamo from a predominantly toroidal field \citep{jacquemin_ide_dynamo_2024,chan_jets_2026}, it may build up gradually, such that the BH spends part of the engine lifetime at $\phit < 1$.\footnote{\citet{wu_maximal_2025} argued that most long GRBs arise far from the MAD state, with $\phit \lesssim 0.05$, where the BH spins up toward $\aeq \approx 1$. However, in collapsars, the jet launches only once $\phit \approx 0.2$, after which the flux grows to the MAD level, $\phi \approx 50$ \citep{issa_magnetically_2025}.}

The spin evolution then depends on how much flux the BH holds. At the jet-launching threshold, the BH does not necessarily spin down, since $\aeq \approx 0.88$ at $\phit = 0.2$. Our fits add the intermediate regime, where the BH spins down slowly toward a moderate $\aeq$: at $\phit = 0.5$, a BH with $\abhzero = 0.5$ stays near $0.5$, whereas one with $\abhzero = 0.9$ spins down only to $\approx 0.73$ after accreting $0.1\,\mbhzero$. Achieving low equilibrium spins, $\astar \lesssim 0.3$, requires $\phit \gtrsim 0.7$, and in the MAD limit, a maximally rotating BH reaches $\astar \lesssim 0.2$ after accreting $\approx 0.2\,\mbhzero$ \citep{jacquemin-ide_collapsar_2023, lowell_rapid_2024}. Thus, the final spin is sensitive to when the disk forms and how quickly it becomes magnetically arrested (\citealp{chan_jets_2026}; see also \citealp{shibata_2024}). The final spin also depends on the disk orientation: bound cocoon material that falls back onto the disk tilts it stochastically, by up to $\sim 60^\circ$, such that the disk and jets become misaligned with the BH spin \citep{gottlieb_black_2022-1}. In this case, the BH accretes angular momentum that is misaligned with its spin, which our calculations do not cover. Neutrino cooling, which becomes important at high $\mdot$, raises the MAD equilibrium spin to $\aeq \approx 0.13$ \citep{issa_collapsar_2026}, so our hot-flow $\aeq$ is a lower limit for collapsars. We leave the recalibration of the fits with neutrino transport to future work.

\subsubsection{Tidal disruption events} 

Classical TDEs by supermassive BHs do not affect the BH spin, since $M_\star/\mbh \ll 1$. In contrast, in micro-TDEs, a $\sim 10\,\msun$ BH in a dense star cluster disrupts a $\sim \msun$ star \citep{perets_micro-tidal_2016,kremer_wind-reprocessed_2023} and accretes up to half of the stellar mass, $M_{0,\rm acc} \lesssim 0.05\,\mbhzero$ per event. Accreting this mass over hours to days corresponds to highly super-Eddington rates, at which the flow is radiatively inefficient and our fits apply. The spin evolution then depends on the magnetic flux. In a MAD, the BH spin evolves toward $\aeq \approx 0.06$. A single event supplies too little mass to reach this equilibrium, but enough for substantial spin-down: a BH with $\abhzero = 0.9$ reaches $\astar \sim 0.5$ after accreting $0.05\,\mbhzero$. In weakly magnetized flows, $\phit \lesssim 0.2$, the equilibrium spin is high, $\aeq \gtrsim 0.88$, but the BH spins up slowly: a non-spinning BH reaches $\astar \approx 0.2$--$0.3$ only after accreting $0.1\,\mbhzero$, i.e., over at least two events.

Jetted micro-TDE models of ultra-long GRBs, such as GRB~250702B \citep{Levan2025}, adopt a rapidly spinning BH, $\astar \approx 0.9$ \citep{kiroglu_jetted_2026}. If the jet is powered by a MAD, BH spin-down may contribute to the decline in jet power over the event, in addition to the declining fallback rate. Conversely, repeated micro-TDEs have been proposed as a dynamical channel for producing rapidly spinning BHs through prograde disk accretion \citep{kritos_compact_2026}. Our fits suggest that building rapidly spinning BHs through repeated micro-TDEs requires predominantly weakly magnetized accretion. Sustained MAD episodes instead drive the BH spin toward a low equilibrium value, limiting this channel's ability to produce high spins. Finally, the debris disk inherits the orbital angular momentum of the disrupted star, which in dense star clusters is randomly oriented with respect to the BH spin. Therefore, the disk is generally misaligned with the BH spin, which our calculations do not cover.

\subsubsection{X-ray binaries}

During binary evolution, an accreting BH may undergo phases of super-Eddington mass transfer \citep{qin_hypercritical_2022, xing_2025}, during which the system may appear as an ultraluminous X-ray source (ULX; \citealt{king_ulx_2023}). The inner flow is then radiatively inefficient, and our fits apply. Binary evolution calculations usually evolve the spin with thin-disk prescriptions \citep{bardeen_kerr_1970, qin_hypercritical_2022,xing_2025}, although it is unclear whether super-Eddington accretion can efficiently spin up the BH \citep{van_son_2020}.

Our fits show that the outcome depends on the horizon magnetic flux. If the jets are weak, $\phit \lesssim 0.2$, the BH spins up at nearly the thin-disk rate toward $\astar \gtrsim 0.9$, so thin-disk prescriptions remain a good approximation. Only powerful jets, $\phit \gtrsim 0.7$, substantially change the spin evolution: their torques spin the BH down toward $\astar \lesssim 0.3$ after it accretes only a fraction of its mass.

Strongly magnetized super-Eddington accretion can therefore keep BH spins low after BH formation, consistent with the predominantly low spins of merging binary BHs \citep{abbott_gwtc-4_2025}. In contrast, producing the high spins inferred for XRBs \citep{draghis_systematically_2024} through accretion would require weakly magnetized mass transfer of a substantial fraction of the BH mass. These spins may instead be natal, or partly reflect the systematic uncertainties in spin measurements discussed by \citet{shashank_measuring_2025}, \citet{tausch_spinning_2026}, and \citet{zdziarski_spins_2026}. Jet-driven spin-down during super-Eddington phases may thus contribute to the tension between the XRB and GW spin distributions \citep{Fishbach2022, zdziarski_spins_2026}.

\subsubsection{Compact-object mergers}

In black hole--neutron star and binary neutron star mergers, the post-merger disk mass is a small fraction of the remnant BH mass, typically $\lesssim 0.2\,\msun$ \citep[e.g.,][]{radice_binary_2018, foucart_remnant_2018}, so $M_{0,\rm acc}/\mbhzero \lesssim 0.1$. If such a disk promptly reaches the MAD state, the remnant spin could decrease by a few tens of percent. However, in the post-merger scenario, BZ jets can launch before the flow becomes magnetically arrested, and spin evolution is at best moderate \citep{Gottlieb_unified_2023}.

\subsubsection{AGN}

For supermassive BHs, our fits apply to the two radiatively inefficient regimes: highly super-Eddington and low-Eddington accretion. The So\l{}tan argument suggests that supermassive BHs gained most of their mass in radiatively efficient quasar phases near the Eddington rate \citep{Soltan1982,yu_tremaine_2002}, which our fits do not cover. However, rapidly growing supermassive BHs, e.g., at high redshift, can accrete a significant fraction of their mass in strongly magnetized, super-Eddington phases \citep{massonneau_2023,kaaz_h-amr_2025}. In such MADs, our fits give $\aeq \approx 0.06$, consistent with the near-zero $\aeq$ that \citet{ricarte_recipes_2023} found for highly super-Eddington MADs \citep[see also][]{kwan_strongly_2026}. Near the Eddington rate, as in luminous quasars, the disk cools radiatively and becomes geometrically thin, so applying our fits requires radiation transport. In this regime, \citet{ricarte_recipes_2023} found $\aeq \approx 0.8$, whereas \citet{lowell_evidence_2025} found that $\aeq$ increases as the MAD thins, reaching $\approx 0.3$ for $h/r \lesssim 0.1$. At low Eddington ratios, \citet{cho_bridging_2026} found, using multizone simulations that connect the horizon to galactic scales, that the BH spin in hot flows evolves on timescales of $\sim 10$\,Gyr, and cautioned against extrapolating spin-down rates from small-domain GRMHD simulations to galactic scales. This caveat applies to our fits as well, which are calibrated on torus-fed flows that span $\lesssim 10^{5}\,\rgzero$ and last $\lesssim 10^{5}\,\rgzero/c$.

\subsection{Application as a subgrid prescription} \label{sec:disc_subgrid}

Galaxy-scale and cosmological simulations increasingly evolve BH spin on the fly \citep{dubois_2014,dubois_2021,talbot_2021,husko_2022,beckmann_black_2025,husko_hybrid_2026}. They typically switch between a thin-disk spin-up rate and a MAD spin-down rate \citep[e.g.,][]{massonneau_2023}.
Eq.~\eqref{eq:spinup_fit} with the coefficients in Table~\ref{tab:fits} provides a continuous prescription for hot, radiatively inefficient flows, spanning weakly magnetized accretion through the MAD regime:
\begin{equation}
    \frac{da_\star}{dt} =s(a_\star,\tilde{\phi})\frac{\mdot}{M_{\rm BH}},
    \qquad
    \frac{dM_{\rm BH}}{dt} =(e_{\rm HD}+e_{\rm EM})\mdot.
\end{equation}
Applying this prescription to cooled flows requires recalibrating the HD and EM fits to account for the modified disk structure, as well as including photon or neutrino energy and angular momentum fluxes through the horizon where relevant. Within the hot-flow regime, the main requirement is a model for $\phit$, which galaxy-scale simulations do not resolve. In practice, $\phit$ must be tied to resolved quantities such as the magnetic flux supplied from large radii or the accretion state. Alternatively, $\phit$ can be constrained by the jet power, $L_{\rm j}$, through the jet efficiency, $\eta_{\rm j} = L_{\rm j} / \mdot \approx -e_{\rm EM}$, and the mass accretion rate, $\mdot$.\footnote{When estimated from $\eta_{\rm j}$, $\phit$ should be capped at $1.2$, the upper limit of our calibration.} For the equilibrium spin of sustained MADs, which saturate at $\phit\approx1.1$--$1.2$ near $\aeq\approx0.05$, we suggest the use of $\phit\approx1.2$, for which Eq.~\eqref{eq:aeq_fit} gives $\aeq\approx0.06$.

\begin{acknowledgments}
D.I. thanks Nicholas Kaaz for fruitful discussions and for valuable contributions to the initial implementation of the moving inner boundary prescription.
J.J. acknowledges support from NSF grants AST-2307983 and NASA grants 80NSSC22K0826, 80NSSC24K0941, and 80NSSC24K1174. 
This research used resources of the Argonne Leadership Computing Facility, a U.S. Department of Energy (DOE) Office of Science user facility at Argonne National Laboratory and is based on research supported by the U.S. DOE Office of Science-Advanced Scientific Computing Research Program, under Contract No. DE-AC02-06CH11357 (NeutronStarRemnants project). The authors acknowledge the MIT Office of Research Computing and Data for providing high performance computing resources on the Engaging cluster that have contributed to the research results reported within this paper.
We acknowledge the use of Claude Opus 5.5 (Anthropic) for coding assistance and simulated peer review. The authors reviewed all AI-generated outputs and take full responsibility for the contents of the final manuscript.
\end{acknowledgments}

\software{\texttt{H-AMR} \citep{hamr_paper}, \texttt{numpy} \citep{numpy_paper}, \texttt{scipy} \citep{scipy_paper}, \texttt{matplotlib} \citep{matplotlib_paper}}

\bibliography{refs}{}
\bibliographystyle{aasjournalv7}

\appendix

\section{Novikov--Thorne disk specific fluxes} \label{app:nt}

An NT disk delivers to the BH the specific energy and angular momentum of circular geodesics at the ISCO \citep{novikov_astrophysics_1973,bardeen_rotating_1972}. In units of $G = c = \mbh = 1$, the ISCO radius is
\begin{equation}
    \begin{split}
        r_{\rm ISCO} &= 3 + Z_2 - {\rm sgn}(\astar) \sqrt{(3 - Z_1)(3 + Z_1 + 2 Z_2)}, \\
        Z_1 &= 1 + \left( 1 - \astar^2 \right)^{1/3} \left[ \left( 1 + \astar \right)^{1/3} + \left( 1 - \astar \right)^{1/3} \right], \\
        Z_2 &= \sqrt{3 \astar^2 + Z_1^2},
    \end{split}
\end{equation}
and the specific energy and angular momentum at $r = r_{\rm ISCO}$ are
\begin{equation}
    e_{\rm NT} = \frac{1 - 2/r + \astar r^{-3/2}}{\sqrt{1 - 3/r + 2 \astar r^{-3/2}}}, \quad
    l_{\rm NT} = \frac{\sqrt{r} \left( 1 - 2 \astar r^{-3/2} + \astar^2 r^{-2} \right)}{\sqrt{1 - 3/r + 2 \astar r^{-3/2}}},
\end{equation}
where $\astar < 0$ corresponds to a disk counter-rotating with respect to the BH. With these fluxes, $s_{\rm NT} = l_{\rm NT} - 2\astar e_{\rm NT}$ vanishes at $\astar = 1$ \citep{bardeen_kerr_1970}.

\section{Fitting the simulation data} \label{app:fits}
\setcounter{equation}{0}
The simulations differ in duration, and each one samples a different part of $(\astar, \phit)$ space. MADs stay near the saturation level, and some of them approach the equilibrium spin. Transient MADs pass through intermediate $\phit$, and SANEs remain at low $\phit$. If every snapshot were fitted as an independent measurement, the accretion states that lasted longest would be overweighted. Neighboring snapshots of a single simulation are also strongly autocorrelated. We therefore reduce each run to a set of approximately independent measurements before fitting. For each simulation we extract time series of the following horizon quantities: the rest-mass accretion rate $\mdot$, the energy and angular-momentum fluxes $\dot E_X$ and $\dot J_X$ ($X=\{\rm HD, EM\}$; corrected for floors as described in Sec.~\ref{sec:methods}), the magnetic flux $\Phi_{\rm H}$, the spin $\astar$ and the current-to-initial BH mass ratio $m$. All fluxes are evaluated on the first radial shell outside the horizon. From these we form the normalized flux $\phi = \sqrt{4\pi}\,\Phi_{\rm H}^{\rm HL}/\left(m \sqrt{\langle\mdot\rangle}\right)$ [Eq.~\eqref{eq:phi}], where $\langle\mdot\rangle$ is a running mean over $10^3\,\rgzero/c$.

The window length is set by the integrated autocorrelation time of $\mdot$, $\tau_{\rm int}=1+2\sum_k\rho(k)$. We first remove slow trends by subtracting a running mean of width $5\times 10^3 \,\rgzero/c$. We then compute $\rho(k)$ by FFT and truncate the sum with an automatic Sokal window, taking the smallest lag $N$ with $N\ge5\,\tau_{\rm int}(N)$. Each run is divided into consecutive windows of at least $n_{\min}=\max(3\tau_{\rm int}, 8)$ samples. A window is extended, up to $5n_{\min}$ samples, as long as $\astar$ and $\phi$ stay close to their values at its start: $|\Delta\astar|<0.02$ and $|\Delta\ln\phi|<0.1$.

For each window we compute the specific energy and angular momentum fluxes as ratios of time-integrated fluxes,
\begin{equation}
e_X = \frac{\int \dot E_X\,dt}{\int \mdot\,dt}, \qquad
l_X = \frac{\int (\dot J_X/m)\,dt}{\int \mdot\,dt},
\end{equation}
which equals the $\mdot$-weighted mean of the instantaneous ratios. The window position $(a_{\star,w}, \phi_w)$ is the $\mdot$-weighted mean spin together with $\phi$ evaluated from the window-averaged $\Phi_{\rm H}/m$ and $\mdot$. Statistical errors come from a delete-one-block jackknife over four contiguous sub-blocks of the window.

\begin{figure}
    \centering
    \includegraphics[width=0.7\linewidth]{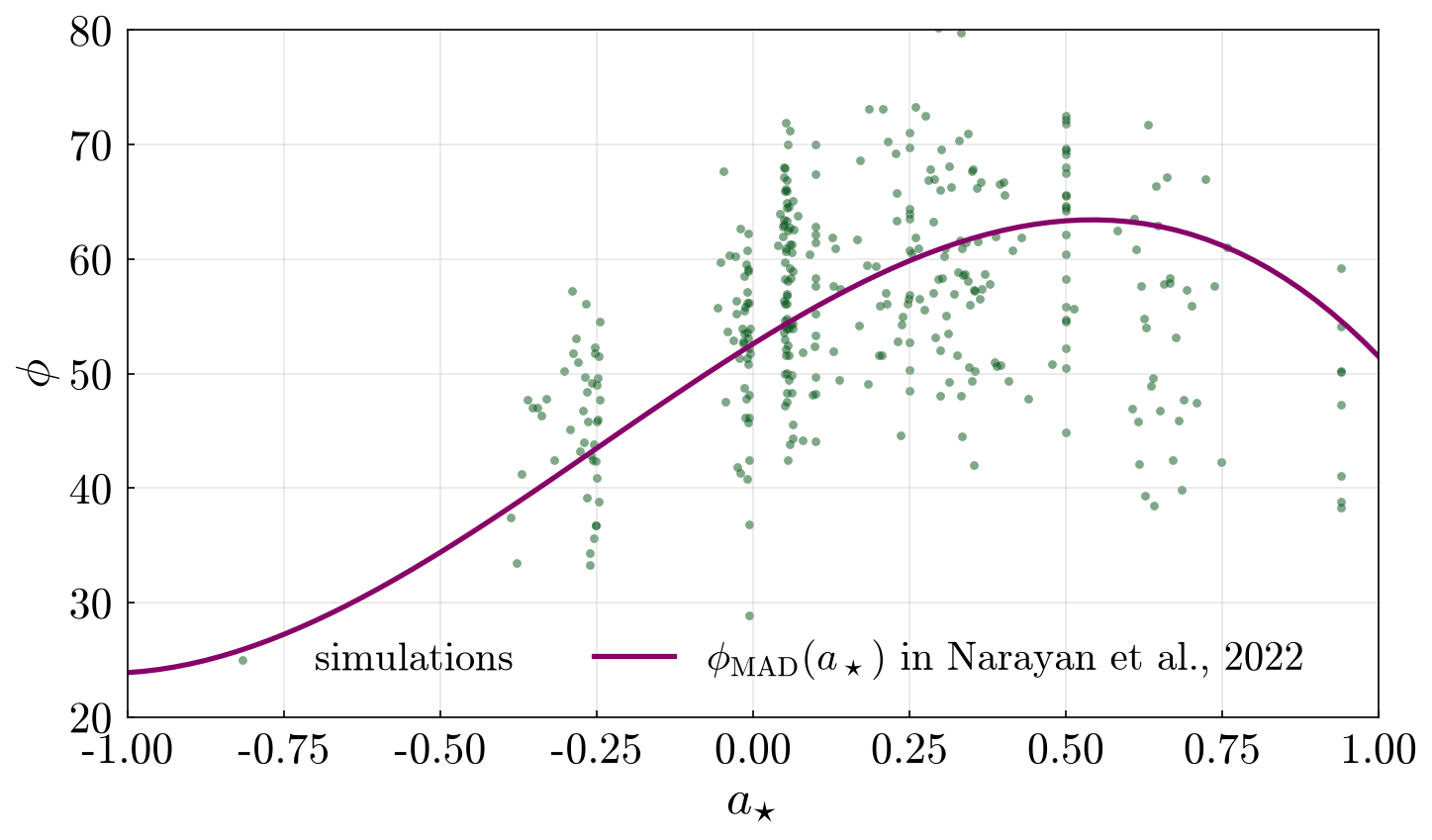}
    \caption{Dimensionless horizon magnetic flux, $\phi$, measured in MAD-state simulation snapshots as a function of BH spin. The solid curve shows the adopted saturation-flux prescription, $\phi_{\rm MAD}$, from \citet{narayan_jets_2022}.}
    \label{fig:phi-mad-fit}
\end{figure}

For the fits we use $\phit=\phi/\phi_{\rm MAD}(\astar)$, where $\phi_{\rm MAD}(\astar)$ is the saturation flux of \citet{narayan_jets_2022}.   Fig.~\ref{fig:phi-mad-fit} compares the horizon flux in our MAD-state snapshots with this fit: near $\astar \approx 0.05$, our MADs saturate at $\phit \approx 1.1$--$1.2$, which places sustained MADs at $\phit > 1$. We keep only windows that start after $5\times 10^3\,\rgzero/c$, to exclude the build-up phase, and that have $\phit \le 1.2$. These windows are binned onto a grid in $(\astar,\phit)$ with cell widths $\Delta\astar\approx0.085$ and $\Delta\phit\approx0.16$. Within each cell, the windows of a single simulation are combined by inverse-variance weighting. The resulting variance is multiplied by $\max(1, N_{\rm win}/10)$, so that one long run contributes at most the weight of ten windows to any cell.

Different simulations that fall in the same cell scatter more than their statistical errors allow. We account for this with a systematic error for each component of the specific energy and angular momentum flux, $\sigma_{\rm sys}^2=\tau_0^2+(f\,|\mu|)^2$, where $\mu$ is the cell mean. The parameters $\tau_0$ and $f$ are fitted by restricted maximum likelihood to all cells that contain at least two simulations. Finally, the per-simulation values in each cell are combined by inverse-variance weighting, with the statistical and systematic variances added. Where residual scatter remains ($\chi^2/{\rm dof}>1$), the cell error is increased by the Birge ratio. This gives one value and one error per quantity per $(\astar,\phit)$ cell. We fit the functional forms in Table~\ref{tab:fits} to these cell values.

\section{Work done by the time-dependent metric} \label{app:metric_work}
\setcounter{equation}{0}
Energy-momentum conservation, $\nabla_\mu T^\mu_\nu = 0$, can be written in coordinate form as
\begin{equation} \label{eq:emom_conservation}
    \partial_t\left(\sqrt{-g}\,T^t_\nu\right) + \partial_i\left(\sqrt{-g}\,T^i_\nu\right) = \sqrt{-g}\,T^\alpha_\lambda \Gamma^\lambda_{\nu\alpha} = \frac{1}{2}\sqrt{-g}\,T^{\alpha\lambda}\,\partial_\nu g_{\alpha\lambda},
\end{equation}
where the second equality follows from the symmetry of $T^{\alpha\lambda}$. For $\nu = \varphi$, the right-hand side vanishes because the metric is axisymmetric, including after the metric updates. For $\nu = t$, it vanishes for a stationary metric, but not when $m$ and $a$ depend on time.

We integrate Eq.~\eqref{eq:emom_conservation} over the volume between the surface at which we measure the horizon fluxes (Sec.~\ref{sec:methods}) and the outer boundary of the grid, $r_{\rm out}$, with $d^3x = dr\,d\theta\,d\varphi$, and neglect the motion of the inner surface. The energy and angular momentum of the gas, $E_{\rm gas} = -\int \sqrt{-g}\,T^t_t\,d^3x$ and $L_{\rm gas} = \int \sqrt{-g}\,T^t_\varphi\,d^3x$, then evolve as
\begin{equation} \label{eq:gas_EL_evolution}
    \begin{split}
        \frac{dE_{\rm gas}}{dt} &= -\frac{d\mbh}{dt} - F_{E,\rm out} + W, \\
        \frac{dL_{\rm gas}}{dt} &= -\frac{dJ}{dt} - F_{L,\rm out},
    \end{split}
\end{equation}
where $d\mbh/dt$ and $dJ/dt$ are given by Eqs.~\eqref{eq:dMdt} and \eqref{eq:dJdt}, $F_{E,\rm out} = -\int \sqrt{-g}\,T^r_t\,d\theta\,d\varphi$ and $F_{L,\rm out} = \int \sqrt{-g}\,T^r_\varphi\,d\theta\,d\varphi$ are the energy and angular momentum outflow rates through $r = r_{\rm out}$, and
\begin{equation} \label{eq:metric_work}
    W = -\frac{1}{2}\int \sqrt{-g}\,T^{\alpha\lambda}\,\partial_t g_{\alpha\lambda}\,d^3x = -\frac{1}{2}\int \sqrt{-g}\,T^{\alpha\lambda}\left(\frac{\partial g_{\alpha\lambda}}{\partial m}\frac{dm}{dt} + \frac{\partial g_{\alpha\lambda}}{\partial a}\frac{da}{dt}\right) d^3x
\end{equation}
is the rate of work done on the gas by the time-dependent metric. The total energy and angular momentum of the gas and the BH therefore satisfy
\begin{equation} \label{eq:total_EL_evolution}
    \frac{d}{dt}\left(E_{\rm gas} + \mbh\right) = -F_{E,\rm out} + W, \qquad \frac{d}{dt}\left(L_{\rm gas} + J\right) = -F_{L,\rm out}.
\end{equation}
  
\begin{figure}
    \centering
    \includegraphics[width=0.7\linewidth]{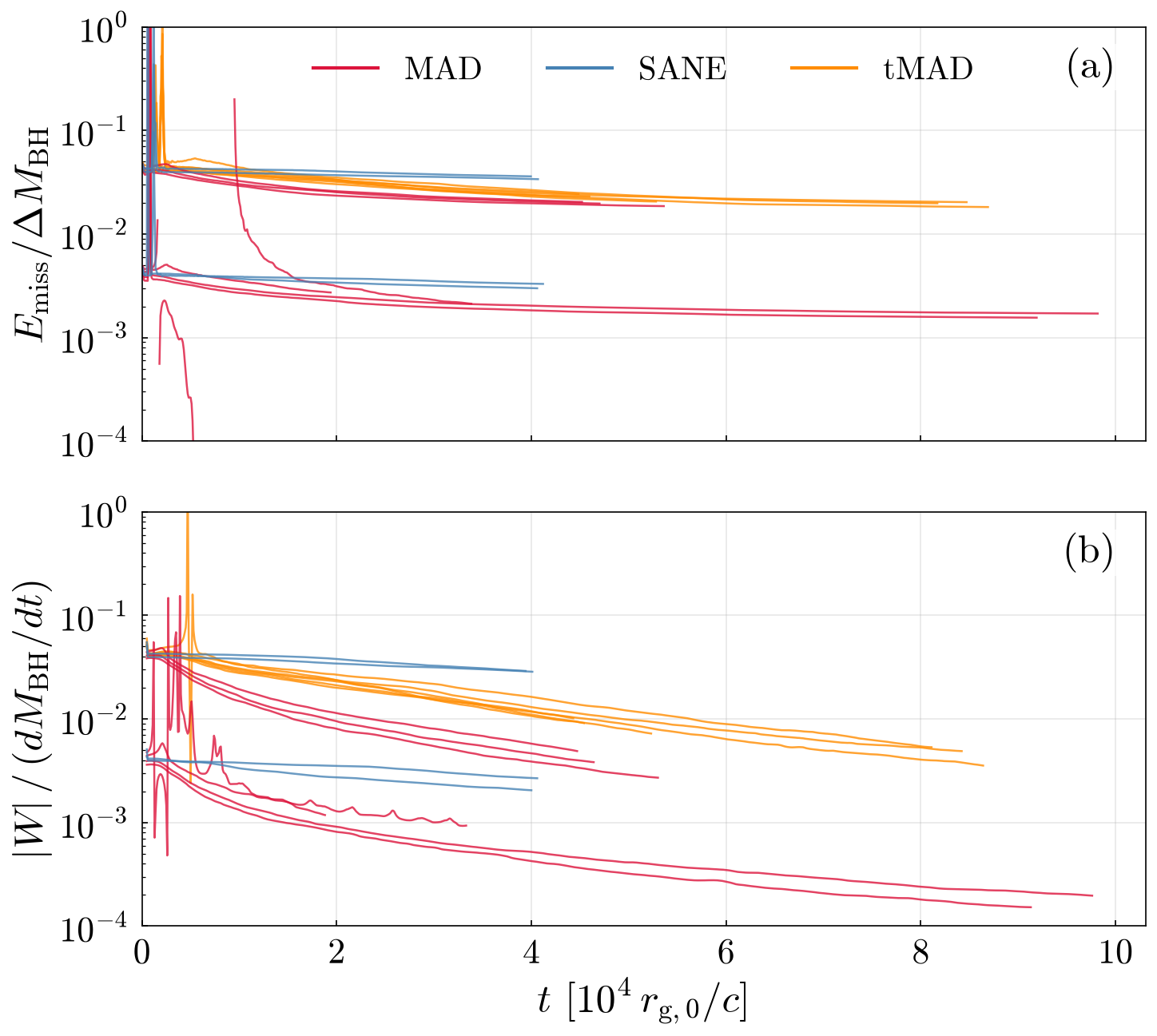}
    \caption{Energy omitted by neglecting the work done by the time-dependent metric in all simulations with evolving BHs. {\bf (a)}: cumulative omitted energy, $|E_{\rm miss}| = |\int_0^t W\,dt^{\prime}|$, in units of the BH mass-energy growth, $\Delta \mbh = \mbh - \mbhzero$. {\bf (b)}: ratio of the instantaneous rate of work done on the gas by the metric, $|W|$, to the BH mass-energy growth rate, $d\mbh/dt$. Colors mark the accretion state. The upper and lower groups of curves correspond to $M_{\rm torus}/\mbhzero \approx 4$--$6$ and $\approx 0.35$--$0.56$, respectively. The spikes at early times occur where $\Delta\mbh$ or $d\mbh/dt$ pass through zero.}
    \label{fig:missing-work}
\end{figure}

In our scheme the metric changes in discrete steps, so that $\partial_t g_{\alpha\lambda} = \sum_n \Delta g^n_{\alpha\lambda}\,\delta(t - t_n)$, where $\Delta g^n_{\alpha\lambda}$ is the change of the metric at the update at $t_n$. Because we hold the vector of conserved variables, $\mathcal{U} = \left[ \sqrt{-g}\rho u^t, \sqrt{-g} T^t_t, \sqrt{-g} T^t_i, \sqrt{-g} B^i \right]$, fixed across each update (Sec.~\ref{sec:methods}), the energy $\Delta E_W^n = -\frac{1}{2}\int \sqrt{-g}\,T^{\alpha\lambda}\,\Delta g^n_{\alpha\lambda}\,d^3x$, evaluated at $t_n$, is not added to the gas, and the scheme conserves $E_{\rm gas} + \mbh$ up to the boundary fluxes, i.e., it solves Eq.~\eqref{eq:total_EL_evolution} with $W = 0$. The angular momentum equation is unaffected. The neglected energy accumulated over a simulation is $E_{\rm miss} = \sum_n \Delta E_W^n = \int W\,dt$, which we compare with the change in the BH mass, $\Delta\mbh$.

The magnitude of $W$ can be estimated in the Newtonian limit, $r \gg \rgzero$, for slowly moving gas. There, the dominant term in Eq.~\eqref{eq:metric_work} is $T^{tt}\,\partial g_{tt}/\partial m \approx 2\rho/r$, so that $W \approx -(dm/dt)\int \rho\, r^{-1}\,dV$. This is the change in the gravitational binding energy of the gas as the BH mass grows: for $dm/dt > 0$, the gas becomes more bound, and $W < 0$. Relative to the BH growth rate, $|W|/(d\mbh/dt) \approx M_{\rm gas}\langle \rgzero/r \rangle_\rho/\mbhzero$, where $M_{\rm gas}$ is the gas mass on the grid and $\langle\cdot\rangle_\rho$ denotes a density-weighted average. The terms proportional to $da/dt$ (from $g_{t\varphi}$ and $g_{\varphi\varphi}$) are suppressed at large radii by additional factors of $\lesssim (\rgzero/r)^{3/2}$, since $T^{t\varphi} \sim \rho\,\Omega$ and $T^{\varphi\varphi} \sim \rho\,\Omega^2$, where $\Omega \approx r^{-3/2}$ is the angular velocity of the gas. For the massive tori, $M_{\rm gas} \approx 5\,\mbhzero$, the measured $|W|/(d\mbh/dt) \approx 0.04$ (Fig.~\ref{fig:missing-work}b) corresponds to $\langle \rgzero/r \rangle_\rho \approx 0.01$, i.e., to gas located at an effective radius of $\sim 100\,\rgzero$.

Figure~\ref{fig:missing-work} shows the estimated work omitted during the metric updates in all simulations with evolving BHs. In the massive tori, $M_{\rm torus}/\mbhzero \approx 4$--$6$, the omitted work amounts to $\approx 2$--$4\%$ of the total mass-energy growth of the BH, $\Delta \mbh$, by the end of the simulations. With $\Delta\mbh \approx 0.2$--$0.75\,\mbhzero$ [Fig.~\ref{fig:time-evolution-plot}(c)], this corresponds to $|E_{\rm miss}| \approx 0.01$--$0.02\,\mbhzero c^2$. In the low-mass tori, $M_{\rm torus}/\mbhzero \approx 0.35$--$0.56$, it is $\lesssim 0.5\%$ at late times, consistent with $|W|/(d\mbh/dt) \propto M_{\rm gas}$. The ratio is larger at early times only where $\Delta\mbh$ passes through zero, e.g., in the MAD model with $\abhzero = 0.9$, in which the BH initially loses mass-energy to the jet. Therefore, the omitted work is a small correction to the energy budget and does not enter the angular momentum budget.

\section{Identification of the MAD state} \label{app:mad_state}
\setcounter{equation}{0}

\begin{figure}
    \centering
    \includegraphics[width=1\linewidth]{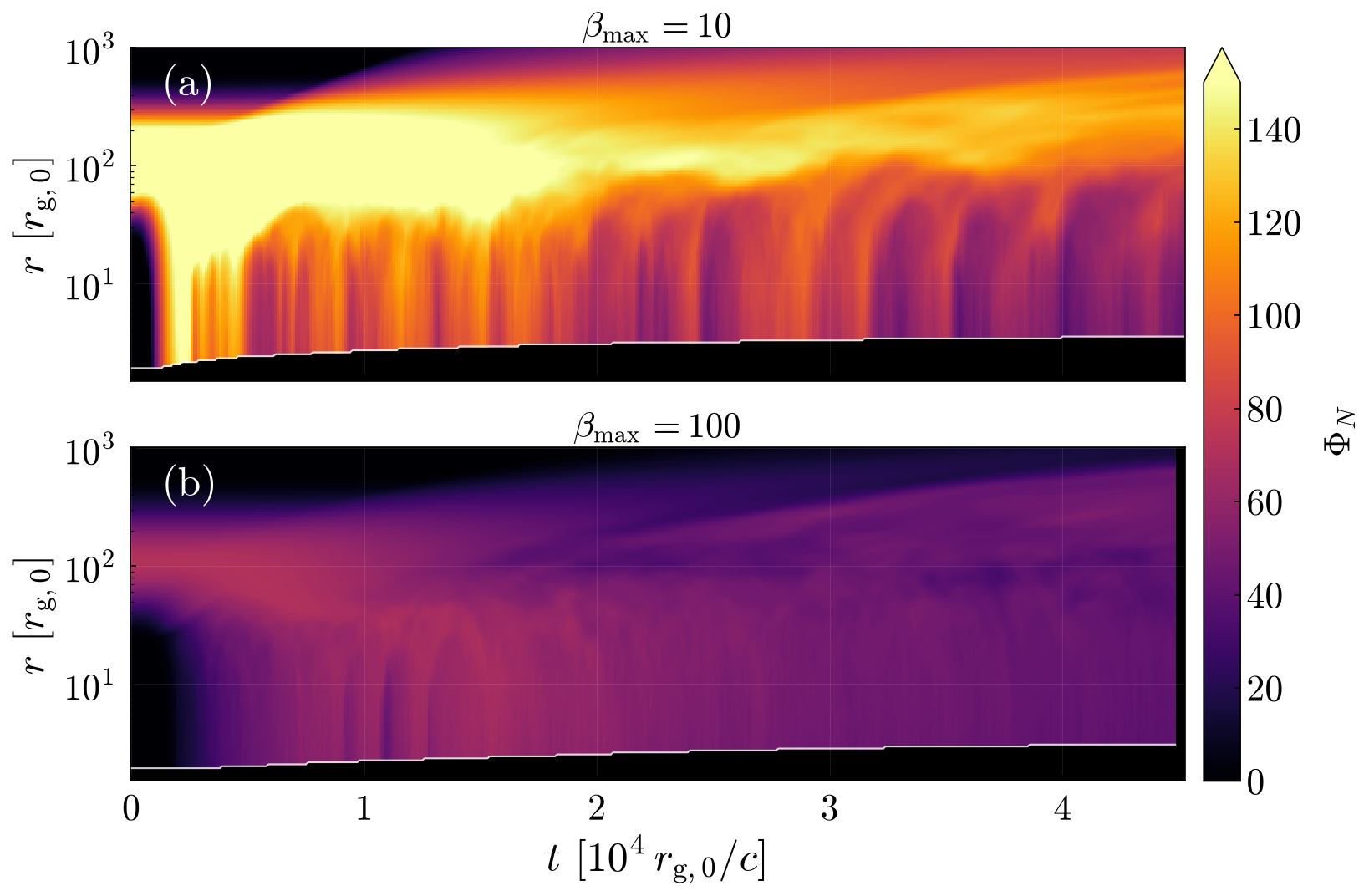}
    \caption{Enclosed magnetic flux threading the northern hemisphere in code units as a function of radius and time for two large-loop models with $\abhzero = 0.5$ and $\beta_{\max}=10$ (a) and $100$ (b). The dark region at the bottom marks the growing event horizon. Panel (a) shows the model that remains in the MAD state throughout the simulation runtime; panel (b) shows the model with the weaker initial field, which reaches the MAD state only transiently. The persistent flux eruptions in (a) show that its MAD state is sustained.}
    \label{fig:mad-comparison}
\end{figure}

Some of the large-loop models reach the MAD state only transiently. Because $\phi \approx \phi_{\rm MAD}$ alone does not establish the MAD state, we also check whether the disks exhibit flux eruptions. Figure~\ref{fig:mad-comparison} shows the magnetic flux threading the northern hemisphere, $\Phi_N$, as a function of radius and time for two models. Flux eruptions appear as vertical stripes, in which magnetic flux is expelled from the BH and the flux at small radii drops. In the model with $\beta_{\max} = 10$, the eruptions persist until the end of the simulation. In the model with $\beta_{\max} = 100$, they occur only at $t \approx (0.5$--$2)\times 10^4\,\rgzero/c$ and cease once the disk leaves the MAD state, although the horizon still holds a substantial fraction of the initial flux.

\end{document}